\documentclass{amsart}[10pt]

\usepackage{amscd,amssymb,latexsym,url,verbatim,graphicx,color,listings}
\usepackage{tikz,tikz-cd}
\usetikzlibrary{decorations.pathmorphing}

\usepackage{cases,amsmath}

\usepackage{txfonts}

\usepackage{tikz,tikz-cd}

\usepackage[dvips]{epsfig}

\title[Combinatorial topology of Weak Symmetry Breaking]
{Combinatorial topology of the standard chromatic subdivision and Weak Symmetry Breaking
for 6 processes}
\author{Dmitry N. Kozlov}

\address{Department of Mathematics, University of Bremen, 28334
  Bremen, Federal Republic of Germany}

\email{dfk@math.uni-bremen.de}


\thanks{We would like to thank Mauruce Herlihy for useful discussions
as well as the anonymous referee for multiple suggestions which have improved
the presentation.}

\newtheorem{theorem}{Theorem}[section]
\newtheorem{df}[theorem]{Definition}
\newtheorem{thm}[theorem]{Theorem} 
\newtheorem{prop}[theorem]{Proposition}
\newtheorem{lm}[theorem]{Lemma}

\newtheorem{rem}[theorem]{Remark}

 \newcommand{\nin}{\noindent}
\newcommand{\pr}{\nin{\bf Proof.} }

\newcommand{\ca}{{\mathcal A}}

\newcommand{\cf}{{\mathcal F}}

\newcommand{\cm}{{\mathcal M}}
\newcommand{\cn}{{\mathcal N}}

\newcommand{\cw}{{\mathcal W}}
\newcommand{\cv}{{\mathcal V}}

\newcommand{\da}{\Delta}

\newcommand{\dl}{\text{\rm dl}\,}

\newcommand{\ex}{\mathcal V}

\newcommand{\mych}{\chi}
\newcommand{\ra}{\rightarrow}

\newcommand{\sm}{\setminus}
\newcommand{\supp}{\text{\rm supp}\,}

\newcommand{\ti}{\tilde}
\newcommand{\wti}{\widetilde}

\newcommand{\ab}{\allowbreak}
\newcommand{\miff}{if and only if }
\newcommand{\kom}{\chi^2(\da^5)}
\newcommand{\lev}{\text{\rm level}\,}
\newcommand{\s}{\,|\,}
\newcommand{\ds}{\,\,\|\,\,}

\newcommand{\ori}{O}
\newcommand{\sig}{R}
\newcommand{\vs}{ $\,$ \vskip5pt\nin}
\newcommand{\car}{\text{\rm carrier}\,}
\newcommand{\myqed}{\qed\vskip5pt}
\newcommand{\flip}{\cf}
\newcommand{\gr}{\Gamma}
\newcommand{\mra}{\longrightarrow}
\newcommand{\mran}{\stackrel{n}{\mra}}
\newcommand{\sra}{\rightsquigarrow}

\newcommand{\st}{\text{\rm St}\,}
\newcommand{\pred}{\text{\rm Pred}\,}
\newcommand{\match}{\text{\bf Match}\,}
\newcommand{\matcht}{\text{\bf Match}}
\newcommand{\exsimp}{{\bf ExSimp}}
\newcommand{\msb}{sb}

\numberwithin{equation}{section}
\numberwithin{figure}{section}
\numberwithin{table}{section}

\begin{document}

\begin{abstract}
In this paper we study a family of discrete configuration spaces, the
so-called {\it protocol complexes}, which are of utmost importance in
theoretical distributed computing. Specifically, we consider questions
of the existance of compliant binary labelings on the vertices of
iterated standard chromatic subdivisions of an~$n$-simplex. The
existance of such labelings is equivalent to the existance of
distributed protocols solving Weak Symmetry Breaking task in the
standard computational model.

As a~part of our formal model, we introduce function $\msb(n)$,
defined for natural numbers $n$, called the {\it symmetry breaking}
function. From the geometric point of view $\msb(n)$ denotes the
minimal number of iterations of the standard chromatic subdivision of
an $(n-1)$-simplex, which is needed for the compliant binary labeling
to exist. From the point of view of distributed computing, the function
$\msb(n)$ measures the minimal number of rounds in a protocol solving
the Weak Symmetry Breaking task.

In addition to the development of combinatorial topology, which is
applicable in a~broader context, our main contribution is the proof of
new bounds for the function $\msb(n)$. Accordingly, the bulk of the
paper is taken up by in-depth analysis of the structure of adjacency
graph on the set of $n$-simplices in iterated standard chromatic
subdivision of an~$n$-simplex. On the algorithmic side, we provide the
first distributed protocol solving Weak Symmetry Breaking task in the
layered immediate snapshot computational model for some number of
processes. 

It is well known, that the smallest number of processes for which Weak
Symmetry Breaking task is solvable is~$6$. Based on our analysis, we
are able to find a~very fast explicit protocol, solving the Weak
Symmetry Breaking for $6$ processes using only $3$
rounds. Furthermore, we show that no protocol can solve Weak Symmetry
Breaking in fewer than 2 rounds.
\end{abstract}

\maketitle

\tableofcontents

\section{Solvability of Weak Symmetry Breaking}

\subsection{Weak Symmetry Breaking as a standard distributed task}
\vs Let $n$ be an integer, such that $n\geq 2$. The Weak Symmetry
Breaking task for $n$ processes is an~inputless task, where the
possible outputs are $0$ and $1$. A~distributed protocol is said to
solve the Weak Symmetry Breaking task if in any execution without
failed processes, there exists at least one process which has value
$0$ as well as at least one process which has value $1$.

In the classical setting, the processes know their id's, and are
allowed to compare them. It is however not allowed that any other
information about id's is used. The protocols with this property are
called {\it comparison-based}.\footnote{Alternative terminology {\it
    rank-symmetric} is also used in the literature.} In practice this
means that behavior of each process only depends on the relative
position of its id among the id's of the processes it witnesses and
not on its actual numerical value.  As a~special case, we note that
each process must output the same value in case he does not witness
other processes at all. Weak Symmetry breaking is a standard task in
theoretical distributed computing, and its solvability in the standard
computational models is a~sophisticated question which has been
extensively studied.

For the rest of this paper we shall fix the computational model to be
the layered immediate snapshot model, see~\cite{HKR}. In this model
the processes use two atomic operations being performed on shared
memory. These operations are: {\it write} into the register assigned
to that process, and {\it snapshot read}, which reads entire memory in
one atomic step. Furthermore, it is assumed that the executions are
well-structured in the sense that they must satisfy the two following
conditions. First, it is only allowed that at each time a~group of
processes gets active, these processes perform a~write operation
together, and then they perform a~snapshot read operation together; no
other interleaving in time of the write and read operations is
permitted.  Such executions are called {\it immediate snapshot}
executions. Second, each execution can be broken up in rounds, where
in every round each non-faulty process gets activated precisely once.

Even though this computational model is seemingly quite restrictive,
it has been proved, see, e.g., \cite{HKR}, that most of the commonly
used shared memory computational models are equivalent, in the sense
of which of the tasks are solvable, to this one.  As we will see, this
model has the major advantage that the protocol complexes have a
comparatively simple topological structure.

\subsection{Previous work}
\vs Several groups of researchers have studied the solvability of the
Weak Symmetry Breaking by means of comparison-based layered immediate
snapshot protocols. Due to primarily work of Casta\~neda and Rajsbaum,
\cite{CR0,CR1,CR2}, it is known that the Weak Symmetry Breaking is
solvable if and only if the number of processes is not a~prime power;
see also \cite{AP} for a~counting-based argument for the
impossibility part. This makes $n=6$ the smallest number of processes
for which this task is solvable.

The combinatorial structures arising in related questions on
subdivisions of simplex paths have been studies in
\cite{ACHP,paths}. The specific case $n=6$ has been studied in
\cite{ACHP}, who has proved the existence of the distributed protocol
which solves the Weak Symmetry Breaking task in 17 rounds.

We refer to the classical textbook~\cite{AW}, as well as the more
recent monograph \cite{AE}, for general background on theoretical
distributed computing, and specifically on impossibility results.
A~general reference for topological methods in distributed computing
is~\cite{HKR}. A~general reference for combinatorial topology
is~\cite{book}. Finally, we recommend the survey \cite{IRR} for
background on symmetry breaking tasks.

\subsection{Our results and outline of the paper}
\vs 

\nin Our main mathematical result is the following theorem.

\begin{thm}\label{thm:main-math}
There exists a compliant binary labeling $\lambda$ of the vertices of
$\chi^3(\da^5)$, such that the restriction of $\lambda$ to any $5$-simplex
of $\chi^3(\da^5)$ is surjective.
\end{thm}

\nin
This immediately implies the following theorem in theoretical
distributed computing.

\begin{thm}\label{thm:main-cs}
There exists a~comparison-based layered immediate snapshot distributed 
protocol solving the Weak Symmetry Breaking task for $6$ processes in $3$ 
rounds.
\end{thm}

We also present the associated explicit distributed protocol. Because of the
standard reduction to a mathematical question, see Theorem~\ref{thm:red},
we can derive Theorem~\ref{thm:main-cs} as a~direct corollary of our
main mathematical statement given in Theorem~\ref{thm:main-math}.

%After developing combinatorics of iterated standard chromatic
%subdivisions in Section~\ref{sect:2}, we give the proof of our main
%theorem in Sections~\ref{sect:3} and~\ref{sect:4}. 
Our central method is the in-depth analysis of the simplicial
structure of the second chromatic subdivision of a~simplex. To this
end, we need to develop the apropriate combinatorial language, in
order to formalize and to work with standard chromatic subdivisions.
This is done in Sections~\ref{sect:2} and~\ref{sect:3}, yielding as
a~byproduct the combinatorial framework which is applicable in a~much
bigger generality.

Once the combinatorics is clear, the plan of our proof is as
follows. First, in Section~\ref{sect:4}, we define some specially
designed binary labeling on the vertices of the second chromatic
subdivision of $\da^5$, calling this the {\it initial labeling}. This
labeling is compliant, but it has many 1-monochromatic 5-simplices.
In the next step, we find a~perfect matching on the set of
1-monochromatic 5-simplices, such that whenever two 5-simplices are
matched, they share a~4-simplex.

In Section~\ref{sect:4} we describe the standard matching, which is
easy to define, but which is not perfect. In Section~\ref{sect:5} we
modify this standard matching by connecting unmatched simplices by
augmenting paths.  This is a~standard technique from matching theory,
and it yields a~perfect matching. Once we have a~perfect matching, it
is easy to use that to achieve a~binary compliant labeling of the
vertices of $\chi^3(\da^5)$ without any monochromatic simplices. This
is done in the end of Section~\ref{sect:5}, and it proves our Main
Theorem~\ref{thm:main-math}.

The remaining two sections are dealing with further specific issues.
The explicit protocol for solving Weak Symmetry Breaking task for $6$
processes in $3$ rounds is given in Section~\ref{sect:5}. We finish
with Section~\ref{sect:6}, where we give a~very short argument for
impossibility of solving Weak Symmetry Breaking task in $1$ round for
any number of processes.

While the focus of this paper has been on the case of 6 processes, it
is possible to refine our analysis to a~much more general setting,
yielding new bounds for the symmetry breaking function, as well as
producing distributed protocols for other numbers of processes.
However, there is a~number of technical issues which need to be
resolved, and there is a~number of new ideas which need to be
introduced, since giving the simplex paths explicitely, as was done in
Table~\ref{table:21paths}, is prohibitive for higher values
of~$n$. For this reason, the extension of the techniques from this
paper to higher values of $n$ will appear in a~separate article,
see~\cite{wsb12,bid}.

\section{Combinatorics of iterated chromatic subdivisions}\label{sect:2}
%\subsection{Basic Notations}
\subsection{Standard chromatic subdivision of a~simplex}
\vs For an arbitrary nonnegative integer $n$, we set
$[n]:=\{0,\dots,n\}$.  We let $\da^n$ denote the standard
$n$-simplex. Note that $\da^n$ has $n+1$ vertices, which we index by
the set $[n]$. For brevity, we shall skip the curly brackets for the
sets consisting of a~single element, and write expressions like $A\cup
x$ and $A\sm x$, rather than $A\cup\{x\}$ and $A\sm\{x\}$.

\begin{df}
Given a set $A$ and an element $x\in A$, we shall call the pair
$\gamma=(A,x)$ a~{\bf node}. We shall say that $x$ is the {\it color}
of this node, and write $C(\gamma)=x$.
\end{df} 
When $W$ is any set of nodes, we set $C(W):=\{C(\gamma)\s\gamma\in
W\}$. The reasons for our terminology will become apparent soon.

\begin{df}\label{df:osp}
An {\bf ordered set partition} $\sigma$ of a set $A$ is an~ordered
tuple $(A_1,\dots,A_t)$ of nonempty subsets of $A$ such that $A$ is
a~disjoint union of $A_1,\dots,A_t$. We shall use the notation
$\sigma=(A_1\s\dots\s A_t)$.
\end{df}

Assume now that we are given an ordered set partition
$\sigma=(A_1\s\dots\s A_t)$ of a~set $A$.

\begin{df} 
 We shall call the set $V(\sigma):= \{(A_1\cup\dots\cup
 A_{i(x)},x)\,|\,x\in A\}$ the {\bf set of nodes of $\sigma$}, where
 $i(x)$ denotes the unique index such that $x\in A_{i(x)}$.
\end{df}

Note that we have $V(\sigma)=\{(A_1\cup\dots\cup A_k,x)\,|\,1\leq
k\leq t,\,\,x\in A_k\}$.

\begin{df}
If $(B,x)$ is a node of an ordered set partition $\sigma$ of a~set
$A$, such that $|B|\geq|A|-1$, then we say that $x$ is {\bf almost
  maximal} with respect to~$\sigma$.
\end{df}

\begin{rem}\label{rem:amax}
For every ordered set partition $\sigma$ of a~set $A$, there exist at
least two elements of $A$, which are almost maximal with respect
to~$\sigma$.
\end{rem}
\nin Indeed, if $\sigma=(A_1\s\dots\s A_t)$, then all elements of
$A_t$ are almost maximal. If $|A_t|=1$, then also all elements of
$A_{t-1}$ are almost maximal.

Given a permutation $\pi=(\pi_0\dots\pi_n)$ of $[n]$, there is
a~natural ordered set partition of the set $[n]$ associated to it,
namely $(\pi_0\s\dots\s \pi_n)$; we shall call it $\hat\pi$.  This
way, we obtain precisely all ordered set partitions consisting of
singletons.

There is a standard mathematical reformulation of the solvability of 
the Weak Symmetry Breaking task which we now proceed to describe. The central 
role in this description is played by the following abstract simplicial 
complex, called {\it standard chromatic subdivision of a simplex}.

\begin{df}\label{df:chi1}
Let $n$ be a nonnegative integer. The abstract simplicial complex
$\mych(\da^n)$ is given as follows:
\begin{itemize}
\item the set of vertices $V(\chi(\da^n))$ consists of all nodes
  $(V,x)$, such that $x\in V\subseteq[n]$;
\item the maximal simplices of $\chi(\da^n)$ are indexed by ordered
  set partitions of $[n]$, where for each ordered set partition
  $\sigma$, its set of vertices is given by the corresponding set of
  nodes $V(\sigma)$;
\item in general $S\subseteq V(\chi(\da^n))$ is a simplex \miff there
  exists a~maximal simplex $\sigma$ such that $S\subseteq V(\sigma)$.
\end{itemize}
\end{df}

The color of the vertex of $\chi(\da^n)$ is the color of its indexing
node, in particular, $C(-)$ can be applied to any subset of
$V(\mych(\da^n))$.  Furthermore, when $\sigma$ is an $n$-simplex of
$\mych(\da^n)$, we set $C(\sigma):=C(V(\sigma))$. Note that
$\chi(\da^n)$ is a~{\it pure} simplicial complex of dimension~$n$,
meaning that all of its maximal simplices have the same dimension.

The simplicial complex $\chi(\da^n)$ has been introduced by 
Herlihy\& Shavit, \cite{HS}, see also a~recent book \cite{HKR}, 
and is a~widely used gadget in theoretical distributed computing. 
It has been proved in \cite{subd} that $\chi(\da^n)$ is a~subdivision 
of an $n$-simplex. A wide generalization of this fact has been proved 
in~\cite{k1,k2}.

%The following notion will be useful for our induction arguments. 
%It measures how far a~particular $n$-simplex is from the middle one.

%\begin{df}
%Give an $n$-simplex $\sigma$ of $\chi(\da^n)$, indexed by
%$(A_1\s\dots\s A_t)$, we call the number $t-1$ the {\bf peripherality} 
%of $\sigma$, and denote it by~$\per(\sigma)$.
%\end{df}

%In particular, the central simplex $([n])$ has peripherality $0$, and
%all other simplices have peripherality between $1$ and $n$. The maximal
%peripherality $n$ is achieved by the $(n+1)!$ $n$-simplices indexed by 
%the permutations of $[n]$. Note, that if two simplices $\sigma$ and 
%$\tau$ share an $(n-1)$-simplex, then their peripheralities differ 
%exactly by~$1$.

\subsection{Iterated chromatic subdivisions and the Weak Symmetry Breaking}\vs

\begin{figure}[hbt]

  %\begin{picture}(0,0)
  %  \special{psfile=#1.pstex}
  %\end{picture}
  \input{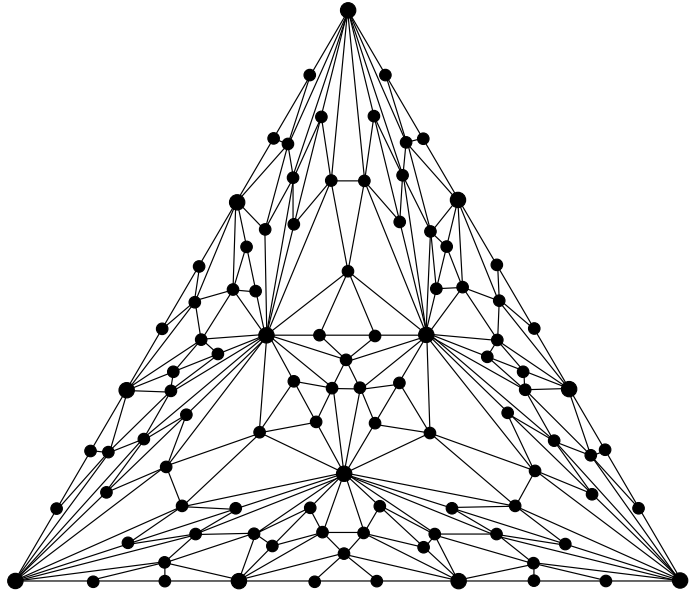}  

\caption{The second iterated standard chromatic subdivision of a~$2$-simplex.}
\label{fig:2}
\end{figure}

The construction from Definition~\ref{df:chi1} can be used to define
chromatic subdivision of an~arbitrary simplicial complex due to the
following simple fact: when restricted to any of its boundary
simplices $\tau$, the standard chromatic subdivision $\mych(\da^n)$ is
naturally isomorphic to $\mych(\da^m)$, where $m$ is the dimension
of~$\tau$. Indeed, given any simplicial complex $K$, we can simply
replace each of its simplices with its standard chromatic subdivision,
these will fit together nicely, and we can call the result the {\it
  standard chromatic subdivision of $K$}. In particular, this means
that we can define {\it iterated chromatic subdivisions}
$\mych^k(\da^n)$.  The simplicial complex $\mych^2(\da^2)$ is shown on
Figure~\ref{fig:2}.

For any finite set $A$ we let $\da^A$ denote the standard simplex
whose vertices are indexed by the elements of $A$. In this setting,
our standard simplex $\da^n$ would be called $\da^{[n]}$. The boundary
simplices are called $\da^I$, for all subsets $I\subset[n]$.

Given two equicardinal subsets $I,J\subset[n]$, the corresponding
boundary simplices $\da^I$ and $\da^J$ have the same dimension, and
any bijection $f:I\ra J$ induces a simplicial isomorphism from $\da^I$
to $\da^J$. Clearly, the construction of the iterated standard
chromatic subdivision will also induce a simplicial isomorphism from
$\chi^d(\da^I)$ to $\chi^d(\da^J)$. We let $\varphi_{I,J}$ denote the
simplicial isomorphism from $\chi^d(\da^I)$ to $\chi^d(\da^J)$ induced
by the unique order-preserving bijection from $I$ to~$J$.

\begin{df} \label{df:cmp}
A binary labeling $\lambda:V(\mych^d(\da^n))\ra\{0,1\}$ is called {\bf
  compliant} if for all $I,J\subset[n]$, such that $|I|=|J|$, 
and all vertices $v\in V(\mych^d(\da^I))$, we have 
\[\lambda(\varphi_{I,J}(v))=\lambda(v).\]
\end{df}
\nin In simple terms, this means that the restriction of the labeling
to $\mych^d(\da^I)$ will only depend on the cardinality of the ordered
set~$I$.

The following well-known statement,
see~\cite{ACHP,CR0,CR1,CR2,HKR,HS,paths}, is a~useful reformulation of
the solvability of the Weak Symmetry Breaking in the layered immediate
snapshot model in purely mathematical terms.

\begin{thm}\label{thm:red}
In the layered immediate snapshot computational model, the Weak
Symmetry Breaking task for $n$ processes is solvable in $d$ rounds if
and only if there exists a~compliant binary labeling
$\lambda:V(\mych^d(\da^{n-1}))\ra\{0,1\}$, such that for every
$(n-1)$-simplex $\sigma\in\mych^d(\da^{n-1})$, the restricted map
$\lambda:V(\sigma)\ra\{0,1\}$ is surjective.
\end{thm}

\nin
Clearly, the reduction in Theorem~\ref{thm:red} means that
Theorem~\ref{thm:main-math} immediately implies
Theorem~\ref{thm:main-cs}.

\subsection{The symmetry breaking function and its estimates}\vs

\nin We are now ready to introduce the main function for our study.

\begin{df}
Assume $n$ is an~arbitrary natural number. We let $\msb(n)$ denote the
minimal number $d$ such that there exists a~compliant binary labeling
$\lambda:V(\mych^d(\da^{n-1}))\ra\{0,1\}$, such that for every
$(n-1)$-simplex $\sigma\in\mych^d(\da^{n-1})$, the restricted map
$\lambda:V(\sigma)\ra\{0,1\}$ is surjective.
If no such $d$ exists, we set $\msb(n):=\infty$.
\end{df}

Note, that once such a~labeling $\lambda$ exists for some $d$, for any
other $d'>d$ it is easy to extend it to a~labeling
$\lambda':V(\mych^{d'}(\da^{n-1}))\ra\{0,1\}$ satisfying the same conditions.

Furthermore, note that Theorem~\ref{thm:red} implies the following
remark.

\begin{rem}
The Weak Symmetry Breaking task for $n$ processes is solvable in the
layered immediate snapshot model if and only if $\msb(n)\neq\infty$.
The actual value $\msb(n)$ is the minimal number of rounds needed for
the distributed protocol to solve this task.
\end{rem}

The Table~\ref{table:sb} summarizes our current knowledge of the
function $\msb(n)$.
\begin{table}[hbt] %\label{table:exv2}
\[\begin{array}{l|l}
\text{Estimate} & \text{Source}\\ \hline
 \msb(n)=\infty \text{ if and only if } n \text{ is a prime power } & \text{\cite{CR0,CR1,CR2}}\\
 \msb(n)=O(n^{q+3}),
 \text{ if } n \text{ is not a~prime power and } & \text{\cite{ACHP}}\\
q \text{ is the largest prime power in
the prime factorization of } n &\\
 \msb(6)\leq 3& \text{Theorem~\ref{thm:main-math}}\\
 \msb(n)\geq 2& \text{Theorem~\ref{thm:sbg2}}
\end{array}\]
%\vspace{0.4cm}
\caption{The known estimates of $\msb(n)$.}
\label{table:sb}
\end{table}

\section{Combinatorial description of the simplicial structure
of standard chromatic subdivisions}

\subsection{Combinatorics of partial ordered set partitions and 
the lower simplices in the standard chromatic subdivision} \vs 
Even though Definition~\ref{df:chi1} provides a~well-defined
simplicial complex, it is somewhat cumbersome to work with, mainly due
to the fact that the lower-dimensional simplices lack a~direct
combinatorial description.  We shall now mitigate this situation by
giving an~alternative combinatorial description of the simplicial
structure of $\chi(\da^n)$, which has been first obtained in
\cite{view}. However, before we can do this, we need some additional
terminology.

\begin{df}
A~{\bf partial ordered set partition} of the set $[n]$ is a~pair of
ordered set partitions of nonempty subsets of $[n]$,
$\sigma=((A_1\s\dots\s A_t), (B_1\s\dots\s B_t))$, which have the same
number of parts, such that for all $1\leq i\leq t$, we have
$B_i\subseteq A_i$.  Given such a~partial ordered set partition
$\sigma$, we introduce the following terminology.
\begin{itemize}
\item The union $A_1\cup\dots\cup A_t$ is called the {\bf carrier set}
of $\sigma$, and is denoted by $\car(\sigma)$.
\item The union $B_1\cup\dots\cup B_t$ is called the {\bf color set}
of $\sigma$, and is denoted by $C(\sigma)$.
\item The {\bf dimension} of $\sigma$ is defined to be $|C(\sigma)|-1$, 
and is denoted $\dim\sigma$.
\end{itemize}
\end{df}

When appropriate, we shall also write 
\begin{equation}\label{eq:tf}
\sigma=\begin{array}{|c|c|c|}
\hline
A_1 & \dots & A_t  \\ \hline
B_1 & \dots & B_t  \\ 
\hline
\end{array}, 
\end{equation}
which we shall call the {\it table form} of~$\sigma$.

We note, that both nodes $(A,x)$, for $A\subseteq[n]$, as well as
ordered set partitions of $[n]$ are special cases of partial ordered
set partitions of~$[n]$. Indeed, a~node $(A,x)$, such that
$A\subseteq[n]$ corresponds to the somewhat degenerate partial ordered
set partition of $[n]$  
\[\sigma=\begin{array}{|c|} \hline A \\ \hline
  x \\ \hline
\end{array}.\] 
Whereas an ordered set partition $(A_1\s\dots\s A_t)$ corresponds to
the partial ordered set partition of $[n]$ 
\[\sigma=\begin{array}{|c|c|c|} \hline A_1&\dots&A_t \\ 
\hline A_1&\dots&A_t
  \\ \hline
\end{array},\]
i.e., a~partial ordered set partition $((A_1\s\dots A_t),
(B_1\s\dots\s B_t))$, such that $A_i=B_i$ for all $i$, and
$A_1\cup\dots\cup A_t=[n]$.

Each partial ordered set partition has non-empty color set, 
which in turn is contained in its carrier set. The nodes 
correspond to the partial ordered set partitions with minimal 
color set, consisting of just one element, and ordered set 
partitions correspond to the partial ordered set partitions 
with maximal color set, namely the whole set $[n]$.

\begin{df}\label{df:dl1}
Assume we are given a partial ordered set partition
$\sigma=((A_1\s\dots\s A_t),\linebreak(B_1\s\dots\s B_t))$ of the set $[n]$,
such that $\dim\sigma\geq 1$, and we are also given an element $x\in
C(\sigma)$, say $x\in B_k$, for some $1\leq k\leq t$. To define the
{\bf deletion of $x$ from $\sigma$} we consider three different cases.
\begin{enumerate}
\item[\text{Case 1.}] If $|B_k|\geq 2$, then the deletion of $x$ from
  $\sigma$ is set to be
\[((A_1\s\dots\s A_t),(B_1\s\dots\s B_{k-1}\s B_k\setminus x\s 
B_{k+1}\s\dots\s B_t)).\]
\item[\text{Case 2.}] If $|B_k|=1$, and $k\leq t-1$, then the deletion
  of $x$ from $\sigma$ is set to be
\[((A_1\s\dots\s A_{k-1}\s A_k\cup A_{k+1}\s\dots\s A_t),
(B_1\s\dots\s B_{k-1}\s B_{k+1}\s\dots\s B_t)).\]
\item[\text{Case 3.}] If $|B_k|=1$, and $k=t$, then the deletion of
  $x$ from $\sigma$ is set to be
\[((A_1\s\dots\s A_{t-1}),(B_1\s\dots\s B_{t-1})).\]
\end{enumerate} 
We denote the deletion of $x$ from $\sigma$ by $\dl(\sigma,x)$.
\end{df}

It is easy to see that $\dl(\sigma,x)$ is again a~partial ordered set
partition of $[n]$. We have $C(\dl(\sigma,x))=C(\sigma)\sm x$,
$\dim\dl(\sigma,x)=\dim\sigma-1$, and
$\car(\dl(\sigma,x))\subseteq\car(\sigma)$.

Definition~\ref{df:dl1} can be generalized as follows. 

\begin{df}\label{df:dl2}
Let $\sigma=((A_1\s \dots\s A_t),\ab(B_1\s\dots\s B_t))$ be a~partial
ordered set partition of the set $[n]$, and assume we are given
a~non-empty set $S\subset C(\sigma)$. Let $i_1<\dots<i_m$ index all
sets $B_i$, such that $B_i\not\subseteq S$, and set $i_0:=0$. For all
$1\leq k\leq m$, we set
\[
\tilde A_k:=A_{i_{k-1}+1}\cup\dots\cup A_{i_k},\text{ and }
\tilde B_k:=B_{i_k}\sm S.\] The obtained partial
ordered set partition $((\ti A_1\s\dots\s\ti A_m),(\ti B_1\s\dots\s\ti
B_m))$ is called {\bf deletion of $S$ from~$\sigma$}, and is denoted
$\dl(\sigma,S)$.
\end{df}

The properties of the deletion of a~part of its color set from
a~partial ordered set partition are summarized in the following
proposition.

\begin{prop}\label{pr:2.7}
Let $\sigma=((A_1\s \dots\s A_t),\ab(B_1\s\dots\s B_t))$ be a~partial
ordered set partition of the set $[n]$. 
\begin{enumerate}
\item[(1)] Assume we have a non-empty subset
  $S\subset C(\sigma)$.\ Then $\dl(\sigma,S)$ is again a~partial
  ordered set partition of the set $[n]$, such that 
\begin{enumerate}
\item[(a)] $C(\dl(\sigma,S))=C(\sigma)\sm S$;
\item[(b)] $\dim\dl(\sigma,S)=\dim\sigma-|S|$;
\item[(c)] $\car(\dl(\sigma,S))\subseteq\car(\sigma)$.
\end{enumerate}
\item[(2)] Assume $S$ and $T$ are disjoint non-empty subsets of
$\supp\sigma$, such that $S\cup T\neq\supp\sigma$. Then, we have
\begin{equation}\label{eq:st}
\dl(\dl(\sigma,S),T)=\dl(\sigma,S\cup T).
\end{equation} 
\end{enumerate}
\end{prop}
\pr All statements are immediate from Definition~\ref{df:dl2}.
\myqed

\nin
Generalizing Definition~\ref{df:osp} we can define the set of nodes of
an arbitrary partial ordered set partition of the set $[n]$.

\begin{df}\label{df:osp2}
Let $\sigma=((A_1\s \dots\s A_t),\ab(B_1\s\dots\s B_t))$ be a~partial
ordered set partition of the set $[n]$. We shall call the set
$V(\sigma)=\{(A_1\cup\dots\cup A_{i(x)},x)\s x\in C(\sigma)\}$, the
{\bf set of nodes} of~$\sigma$, where again $i(x)$ denotes the unique
index such that $x\in B_{i(x)}$.
\end{df}

\nin Note, that $V(\sigma)=\{(A_1\cup\dots\cup A_k,x)\,|\,1\leq k\leq
t,x\in B_k\}$, $|V(\sigma)|=\dim\sigma+1$, and
$C(V(\sigma))=C(\sigma)$. Comparing Definitions~\ref{df:dl2}
and~\ref{df:osp2} yields the identity
\begin{equation}\label{eq:vs2}
V(\sigma)=\{\dl(\sigma,[n]\sm\{x\})\s x\in C(\sigma)\}.
\end{equation} 
Furthermore, when an~ordered set partition is viewed as a partial one,
its set of nodes does not depend on which one of the
Definitions~\ref{df:osp} and~\ref{df:osp2} is used. Crucially, the
set of nodes completely determines any partial ordered set partition
of the set~$[n]$.

\begin{prop}\label{pr:unique} 
Assume $\sigma$ and $\tau$ are both partial ordered set partitions of 
the set $[n]$, such that $V(\sigma)=V(\tau)$, then $\sigma=\tau$.
\end{prop}
\pr Assume $\sigma\neq\tau$. Without loss of generality, we can write
$\sigma=((A_1\s \dots\s A_t),\ab(B_1\s\dots\s B_t))$ and $\tau=((C_1\s
\dots\s C_q),\ab(D_1\s\dots\s D_q))$, such that $t\leq q$. To start
with, the sets of sets $\{A_1,A_1\cup A_2,\dots,A_1\cup\dots\cup
A_t\}$ and $\{C_1,C_1\cup C_2,\dots,C_1\cup\dots\cup C_q\}$ must be
equal. This immediately implies that $t=q$, and that $A_i=C_i$, for
all $i=1,\dots,t$.

Since $C(\sigma)=C(\tau)$, we have $B_1\cup\dots\cup B_t=D_1\cup\dots
\cup D_t$. Let $k$ denote the minimal number such that $B_k\neq D_k$.
Without loss of generality we can assume that we can find an element
$x$, such that $x\in B_k$, $x\not\in D_k$, and $x\in D_m$, for some
$m>k$.  Then, $\sigma$ has a~node $(A_1\cup\dots\cup A_k,x)$, whereas
$\tau$ has a~node $(A_1\cup\dots\cup A_m,x)$. Since $m\neq k$ we
arrive at a~contradiction.  
\myqed

\nin
We are now ready to prove that partial ordered set partitions provide
the right combinatorial language to describe the simplicial structure
of $\chi(\da^n)$.

\begin{prop}\label{pr:chi2}
The nonempty simplices of $\chi(\da^n)$ can be indexed by all partial
ordered set partitions of $[n]$.  This indexing satisfies the following
properties:
\begin{enumerate}
\item[(1)] The dimension of the simplex indexed by
  $\sigma=((A_1\s\dots\s A_t),(B_1\s\dots\s B_t))$ is equal to
  $\dim\sigma$.
\item[(2)] The vertices of the simplex $\sigma$ indexed by
  $((A_1\s\dots\s A_t),(B_1\s\dots\s B_t))$ are indexed by $V(\sigma)$.
\item[(3)] In general, the set of subsimplices of the simplex $\sigma$
  indexed by $((A_1\s\dots\s A_t),\ab(B_1\s\dots\s B_t))$ is precisely
  the set of simplices indexed by partial ordered set partitions from
  the set $\{\dl(\sigma,S)\s S\subset C(\sigma)\}$.
\end{enumerate}
\end{prop}

\pr Assume $W\subseteq V(\chi(\da^n))$, such that $W$ forms a~simplex
in $\mych(\da^n)$. By Definition~\ref{df:chi1} there exists an
$n$-simplex $\sigma$ of $\chi(\da^n)$, such that $W\subseteq
V(\sigma)$. Furthermore, by~\eqref{eq:vs2} we get
$W=\{\dl(\sigma,[n]\sm x)\s x\in C(W)\}$. We now set
\begin{equation}\label{eq:index}
\ti\sigma:=\dl(\sigma,[n]\sm C(W))
\end{equation} 
to be the partial ordered set partition of the set $[n]$ which indexes
the simplex~$W$. First, we note that by Proposition~\ref{pr:2.7}(1)(a)
we have
\begin{equation}\label{eq:h1}
C(\ti\sigma)=[n]\sm([n]\sm C(W))=C(W).
\end{equation} 
Furthermore, we derive 
\begin{multline}\label{eq:ml1}
V(\ti\sigma)=\{\dl(\ti\sigma,C(\ti\sigma)\sm x)\s x\in C(\ti\sigma)\}%=\\
=\{\dl(\ti\sigma,C(W)\sm x )\s x\in C(W)\}=\\
=\{\dl(\dl(\sigma,[n]\sm C(W)),C(W)\sm x )\s x\in C(W)\}%=\\
=\{\dl(\sigma,[n]\sm x )\s x\in C(W)\}=W,
\end{multline}
where the penultimate equality is due to~\eqref{eq:st}.

Assume $\tau$ is another $n$-simplex of $\chi(\da^n)$, such that
$W\subseteq V(\tau)$.  Set $\ti\tau:=\dl(\tau,[n]\sm C(W))$. We have
shown in~\eqref{eq:ml1} that $V(\ti\sigma)=V(\ti\tau)=W$. It follows
from Proposition~\ref{pr:unique} that $\ti\sigma=\ti\tau$, and hence
the partial ordered set partition of the set $[n]$, which indexes $W$
does not depend on the choice of $\sigma$.  

We can now show that the indexing given by~\eqref{eq:index} satisfies
the $3$ properties in the formulation of the proposition. To see
property $(1)$, let $W\subseteq V(\chi(\da^n))$ be the simplex of
$\chi(\da^n)$ indexed by $\ti\sigma$. On one hand, $\dim W=|W|-1$. On
the other hand, picking an $n$-simplex $\sigma$ such that $W\subseteq
V(\sigma)$, we get 
\[\dim(\dl(\sigma,[n]\sm
C(W)))=\dim\sigma-(n+1-|C(W)|)= n-(n+1-|W|)=|W|-1,\] 
hence $\dim W=\dim\ti\sigma$. 

Finally note, that property $(2)$ has been proved by~\eqref{eq:ml1},
whereas property $(3)$ follows directly from~\eqref{eq:st}.
\qed

\subsection{Combinatorial language encoding simplices in iterated 
chromatic subdivisions}
\vs
The following combinatorial concept is the key to describing 
the iterated standard chromatic subdivisions.

\begin{df}\label{df:linked}
Assume we are given a tuple $\sigma=(\sigma_1,\dots,\sigma_d)$ of
partial ordered set partitions of the set $A$. We say that the tuple
$\sigma$ is {\bf linked} if for all $1\leq i\leq d-1$, we have
$C(\sigma_i)=\car(\sigma_{i+1})$.
\end{df}

Assume we have a~partial ordered set partition $\sigma=((A_1\s\dots\s
A_t), (B_1\s\dots\s B_t))$, and $S\subset C(\sigma)$. Let $i$ be the
minimal index, such that $B_i\cup\dots\cup B_t\subseteq S$.  We define
$D(\sigma,S)$ to be $A_i\cup\dots\cup A_t$, if such an index $i$
exists, and we let $D(\sigma,S)$ be empty otherwise.  We can now
generalize Proposition~\ref{pr:chi2}.

\begin{prop}\label{pr:chi3}
The simplices of $\chi^d(\da^n)$ can be indexed by linked $d$-tuples
of partial ordered set partitions of the set $[n]$. This indexing
satisfies following properties:
\begin{enumerate}
\item[(1)] The dimension of the simplex indexed by
  $\sigma=(\sigma_1,\dots,\sigma_t)$ is equal to $\dim\sigma_t$.
\item[(2)] The subsimplices of $\sigma=(\sigma_1,\dots,\sigma_t)$ are
  all tuples $(\dl(\sigma_1,S_1)\dots,\dl(\sigma_t,S_t))$, where
  $S_t\subset C(\sigma_t)$, and $S_i=D(\sigma_{i+1},S_{i+1})$, for all
  $1\leq i\leq t-1$.
\end{enumerate}
\end{prop}
\pr Clearly, each simplex of $\chi^d(\da^n)$ can be obtained by first
choosing a~simplex of $\chi(\da^n)$, then viewing this simplex as
$\da^k$, for some $k\leq n$, then picking the next simplex in $\da^k$,
and so on, repeating $d$ times in total. By Proposition~\ref{pr:chi2}
each next simplex can be indexed by a~partial ordered set partition,
and each time the color set of the previous simplex is the carrier of
the next one. This is exactly the same as requiring for this tuple of
simplices to be linked in the sense of Definition~\ref{df:linked}.
Both (1) and (2) now follow immediately from Proposition~\ref{pr:chi2}
and the definition of the deletion operation.  
\myqed

\nin
Generalizing \eqref{eq:tf}, the index of a simplex in $\chi^d(\da^n)$
can be visualized as an array of subsets:
\[\begin{array}{|c|c|c||c||c|c|c|}
\hline
A_1^1 & \dots & A_{t_1}^1 & \dots & A_1^d & \dots & A_{t_d}^d \\ \hline
B_1^1 & \dots & B_{t_1}^1 & \dots & B_1^d & \dots & B_{t_d}^d \\ 
\hline
\end{array},\]
satisfying $B_1^k\cup\dots\cup B_{t_k}^k=A_1^{k+1}\cup\dots\cup
A_{t_{k+1}}^{k+1}$, for all $1\leq k\leq d-1$.

In particular, the top-dimensional simplices of $\chi^d(\da^n)$ are
indexed by tuples $\alpha=((A_1^1\s\dots\s A_{t_1}^1),\dots,
(A_1^d\s\dots\s A_{t_d}^d))$. We find it practical to use the following
shorthand notation: $\alpha=(A_1^1\s\dots\s A_{t_1}^1\ds\dots\ds
A_1^d\s\dots\s A_{t_d}^d)$.

\subsection{Combinatorics of the pseudomanifold structure of the chromatic
subdivision of a~simplex}
\vs
One of the reasons, why the layered immediate snapshot computational model 
is amenable to detailed analysis is because the corresponding protocol 
complexes have a~useful structure of a~pseudomanifold.

\begin{df}
A pure $n$-dimensional simplicial complex $K$ is called 
a~{\bf pseudomanifold} of dimension $n$ if the following two conditions are 
satisfied:
\begin{enumerate}
\item[(1)] any $(n-1)$-simplex belongs to at most two $n$-simplices;
\item[(2)] any two $n$-simplices $\sigma$ and $\tau$ can be connected by 
a~path of $n$-simplices $\sigma=\sigma_0,\sigma_1,\dots,\sigma_t=\tau$, 
such that for any $k=1,\dots,t$, the $n$-simplices $\sigma_{k-1}$ and 
$\sigma_k$ share an $(n-1)$-simplex.
\end{enumerate}
\end{df}

It is well known, see e.g., \cite{HKR}, that the iterated chromatic
subdivision of an~$n$-simplex is an~$n$-pseudomanifold. We now provide
the combinatorial language for describing how to move between
$n$-simplices of this pseudomanifold.

To start with, consider an $n$-simplex $\sigma$ of $\chi(\da^n)$, say
$\sigma=(A_1\s\dots\s A_t)$.  We set 
\[F(\sigma):=
\begin{cases}
[n]\sm A_t, & \text{ if } |A_t|=1; \\
[n], & \text{ otherwise. }
\end{cases}\]

\begin{df}\label{df:flip}
Given an $n$-simplex $\sigma$ of the simplicial complex $\chi(\da^n)$,
and $x\in F(\sigma)$, we let $\flip(\sigma,x)$ denote the $n$-simplex
of $\chi(\da^n)$ obtained in the following way. Let us say
$\sigma=(A_1\s\dots\s A_t)$, and $x\in A_k$.
\begin{enumerate}
\item[\text{\bf Case 1.}] Assume $|A_k|\geq 2$, then we set 
\[\flip(\sigma,x):=(A_1\s\dots\s A_{k-1}\s x \s A_k\sm x\s A_{k+1}\s\dots\s A_t).\]
\item[\text{\bf Case 2.}] Assume $|A_k|=1$ (that is $A_k= x $) and
  $k<t$. Then we set
\[\flip(\sigma,x):=(A_1\s\dots\s A_{k-1}\s x \cup A_{k+1}\s A_{k+2}\s\dots\s A_t).\]
\end{enumerate}
\end{df}

Note that since $x\in F(\sigma)$, we cannot have the case
$\sigma=(A_1\s\dots\s A_{t-1}\s x)$, hence $\flip(\sigma,x)$ is
well-defined.  We say that $\flip(\sigma,x)$ is obtained from $\sigma$
by a~{\it flip} of $\sigma$ with respect to $x$. In this sense,
$F(\sigma)$ is the set of all colors which can be flipped.  Given any
simplex $\sigma$ we can always flip with respect to all colors except
for at most one color. We can also flip back, so we have
$\flip(\flip(\sigma,x),x)=\sigma$ for all $\sigma$,~$x$.

Furthermore, we remark that for any $x\in F(\sigma)$, the
$n$-simplices $\sigma$ and $\flip(\sigma,x)$ share an $(n-1)$-simplex
$\tau$ given by $\tau=\dl(\sigma,x)=\dl(\flip(\sigma,x),x)$.

Definition~\ref{df:flip} can be generalized to iterated chromatic
subdivisions.  For an~$n$-simplex
$\sigma=(\sigma_1\ds\dots\ds\sigma_d)$ of $\chi^d(\da^n)$ we set
$F(\sigma):=F(\sigma_1)\cup\dots\cup F(\sigma_d)$. Effectively this
means that $F(\sigma)=[n]$, unless
$F(\sigma_1)=\dots=F(\sigma_d)=[n]\sm p$, in which case we have
$F(\sigma)=[n]\sm p$.

\begin{df}\label{df:gflip}
Assume we are given an $n$-simplex $\sigma=(\sigma_1\ds\dots\ds\sigma_d)$
of the simplicial complex $\chi^d(\da^n)$, and $x\in F(\sigma)$. Let
$k$ be the maximal index such that $x\in F(\sigma_k)$, by the
definition of $F(\sigma)$, such $k$ must exist. We define
$\flip(\sigma,x)$ to be the following $d$-tuple of ordered partitions:
\begin{equation}\label{eq:flip}
\flip(\sigma,x):=(\sigma_1\ds\dots\ds\sigma_{k-1}\ds \flip(\sigma_k,x)\ds
\sigma_{k+1}\ds\dots\ds\sigma_n).
\end{equation}
\end{df}

Again, it is easy to see that for any $x\in F(\sigma)$, the $n$-simplices 
$\sigma$ and $\flip(\sigma,x)$ will share an~$(n-1)$-simplex, and that
for all $\sigma$, $x$, we have the identity
\begin{equation}\label{eq:ffx}
\flip(\flip(\sigma,x),x)=\sigma.
\end{equation}

\subsection{Standard chromatic subdivisions and matchings}
\vs
Let us review some basic terminology of the graph theory, more
specifically the matching theory. To start with, recall that a~graph
$G$ is called {\it bipartite} if its set of vertices $V(G)$ can be
represented as a~disjoint union $A\cup B$, such that there are only
edges of the type $(v,w)$, with $v\in A$, $w\in B$. We shall call
$(A,B)$ a~{\it bipartite decomposition}. Note, that such a
decomposition need not be unique.

Classically, a~{\it matching} in a~graph $G$ is a subset $M$ of its
set of edges $E(G)$, such that two different edges from $M$ do not
share vertices. We call edges which belong to $M$ the {\it matching
  edges} and all other edges the {\it non-matching edges}.  A~vertex
$v$ is called {\it matched} if there exists an edge in $M$ having $v$
as an~endpoint. We call the unmatched vertices {\it critical}, which
is consistent with the terminology we used in the previous
sections. A~matching is called {\it perfect} if there are no critical
vertices, otherwise we may call the matching {\it partial}. Clearly,
an existence of a perfect matching implies that the sets $A$ and $B$
have the same cardinality.  A matching is called {\it near-perfect} if
there is exactly one critical vertex.

\begin{figure}[hbt]

  %\begin{picture}(0,0)
  %  \special{psfile=#1.pstex}
  %\end{picture}
  \input{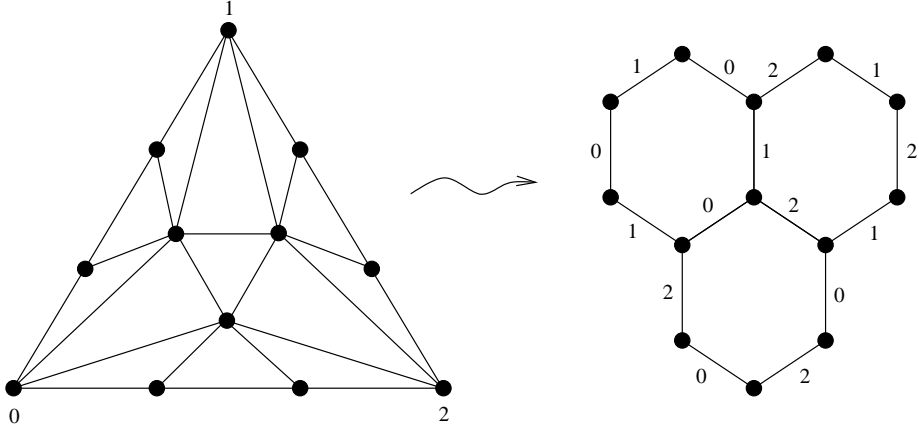}  

\caption{The standard chromatic subdivision of a~$2$-simplex and the corresponding
graph $\gr_2$, with labels showing the colors of the edges.}
\label{fig:1}
\end{figure}

In this paper, the fundamental instance of a graph on which matchings
are constructed is provided by $\gr_n$. This is the graph whose
vertices are all $n$-simplices of $\chi(\da^n)$, and two vertices are
connected by an edge if the corresponding $n$-simplices share
an~$(n-1)$-simplex. Sometimes, we shall abuse our language and call
the vertices of $\gr_n$ simplices. We color the edges of $\gr_n$ as
follows: the edge connecting $\sigma$ with $\tau$ gets the color of
the vertex of $\sigma$ which does not belong to $\sigma\cap\tau$.
This graph has also been studied in~\cite{AC}; an example is shown on
Figure~\ref{fig:1}.

The graph $\gr_n$ is bipartite and the bipartite decomposition is
unique. Recall that $n$-simplices of $\chi(\da^n)$ are indexed by
ordered set partitions $(A_1\s\dots\s A_t)$ of the set~$[n]$. The
bipartite decomposition of $\gr_n$ is then provided by sorting the
$n$-simplices of $\chi(\da^n)$ into two groups according to the parity
of the number~$t$.  For convenience of notations we shall define a
function $\ori:V(\gr_n)\ra\{\pm 1\}$, which we call {\it orientation}
of the simplex, as follows:
\[\ori(A_1\s\dots\s A_t):=\begin{cases}
 1, & \textrm{ if } t \textrm{ is even};\\
-1, & \textrm{ if } t \textrm{ is odd}.
\end{cases}\]
 The matching itself in this context
will mean to group $n$-simplices in pairs, so that in each pair
the $n$-simplices share a~boundary $(n-1)$-simplex. Accordingly, 
we can talk about critical $n$-simplices etc.

In general, for any $d\geq 1$, we let $\gr_n^d$ denote the graph,
whose vertices are the $n$-simplices of $\chi^d(\da^n)$, and two
vertices are connected by an edge if the corresponding $n$-simplices
share an~$(n-1)$-simplex. The edges of $\gr_n^d$ are colored in the
same way as those of~$\gr_n$.

\section{The standard matching for the initial labeling}\label{sect:3}
\subsection{Combinatorics of the second chromatic subdivision of an $n$-simplex}
\vs 
In this paper we will primarily need the combinatorial description
of the simplicial structure of the second chromatic subdivision of an
$n$-simplex.  In this case, the Proposition~\ref{pr:chi3} says that
simplices of $\chi^2(\da^n)$ are indexed by pairs $(\sigma\ds\tau)$ of
partial ordered partitions of $[n]$, such that $C(\sigma)=\car(\tau)$.
The $n$-simplices of $\chi^2(\da^n)$ are simply pairs $(\sigma\ds\tau)$
of $n$-simplices of $\chi(\da^n)$. The vertices of $\chi^2(\da^n)$ are
indexed by pairs of partial ordered set
partitions 
\begin{equation}\label{eq:vert}
v=\begin{array}{|c|c|c||c|} \hline A_1 & \dots & A_t & S
\\ \hline B_1 & \dots & B_t &  x  \\ \hline
\end{array},
\end{equation}
such that $S=B_1\cup\dots\cup B_t$. 

Assume we are given a simplex $\alpha=(\sigma\ds\tau)$ of
$\chi^2(\da^n)$, say $\sigma=((A_1\s\dots\s A_t),\ab(B_1\s\dots\s B_t))$,
and $\tau=((C_1\s\dots\s C_q),(D_1\s\dots\s D_q))$. Assume
furthermore, we are given some subset $S\subset D_1\cup\dots\cup
D_q$. Let $k$ be the minimal index such that $D_k\cup\dots\cup
D_q\subseteq S$, then $\dl(\alpha,S)=(\dl(\sigma,C_k\cup\dots\cup
C_q),\dl(\tau,S))$.

Finally, we remark that for an $n$-simplex $\alpha=(A_1\s\dots\s
A_t\ds B_1\s\dots\s B_q)$, and $x\in[n]$, the flip $\flip(\alpha,x)$
is always defined unless $A_t=B_q= x $, and is explicitly given
by~\eqref{eq:flip}.

\subsection{Description of the initial labeling}
\vs For an arbitrary simplex $\sigma=(\sigma_1\s\dots\s \sigma_d)$ of
$\chi^d(\da^n)$ we set $\supp(\sigma):=\car(\sigma_1)$.

\begin{df}\label{df:int}
Let $v$ be a vertex of $\chi^d(\da^n)$. We say that the vertex $v$ 
is an~{\bf internal} vertex if $\supp(v)=[n]$, otherwise we say that $v$ is 
a~{\bf boundary vertex}.
\end{df}

In the case $d=2$, there is a~handy criterion for deciding whether all
vertices of an $n$-simplex are internal.

\begin{prop}\label{prop:int}
%Let $v$ be a vertex of $\kom$ given as
%\begin{equation}\label{eq:v2}
%v=\begin{array}{|c|c|c||c|}
%\hline
%A_1 & \dots & A_t & S \\ \hline
%B_1 & \dots & B_t &  x  \\ 
%\hline
%\end{array}.
%\end{equation}
%The vertex $v$ is internal if and only if $A_1\cup\dots\cup A_t=[n]$.%
%
Let $\alpha=(A_1\s\dots\s A_k\ds B_1\s\dots\s B_m)$ be an
$n$-simplex of $\chi^2(\da^n)$. Then all vertices of $\alpha$ are internal if
and only if $A_k\cap B_1\neq\emptyset$. 

In general, assume that $A_k\cap B_1=\emptyset$, let $q$ be the
largest index such that $A_k\cap B_i=\emptyset$ for all $1\leq i\leq
q$.  The boundary vertices of $\alpha$ are precisely the vertices with
colors from $B_1\cup\dots\cup B_q$.
\end{prop}
\pr Recall, that since $\alpha$ is an $n$-simplex, we have
$A_1\cup\dots\cup A_k=B_1\cup\dots\cup B_m=[n]$. Pick $x\in[n]$, say
$x\in B_l$. The vertex of $\alpha$, which is colored by $x$ has the
index $(\ti\sigma\ds\ti\tau)$, where $\ti\sigma= \dl(\sigma,B_{l+1}\cup
\dots \cup B_m)$, and $\ti\tau=(B_1\cup\dots\cup B_l\s x)$. This
vertex is internal if and only if $\car(\ti\sigma)=[n]$. By the
definition of the deletion operation, this is the case if and only if
$A_k\sm(B_{l+1}\cup\dots\cup B_m)\neq\emptyset$, i.e.,
\begin{equation}\label{eq:int1}
A_k\cap(B_1\cup\dots\cup B_l)\neq\emptyset.
\end{equation}
Clearly, the fact that \eqref{eq:int1} is true for all $l=1,\dots,m$
is equivalent to the condition $A_k\cap B_1\neq\emptyset$, and in
general \eqref{eq:int1} yields the description of all boundary
vertices of~$\alpha$.  
\myqed

We are now in a~position to describe the initial labeling of the
vertices of $\chi^2(\da^5)$
\[I:V(\chi^2(\da^5))\longrightarrow\{0,1\}.\]
Before we do this, we would like to designate certain boundary vertices 
of $\kom$ as exceptional.

\begin{df}\label{df:exv}
The set $\ex$ consists of all vertices of $\kom$ listed in
Table~\ref{table:exv2}. The vertices in $\ex$ are called {\bf
  exceptional vertices}, while all other boundary vertices are called {\bf
  regular vertices}.
\end{df}

\begin{table}[hbt] %\label{table:exv2}
\[\begin{array}{llll}
\begin{array}{|c||c|}
\hline
a & a \\ \hline
a & a \\ 
\hline
\end{array} &&& \text{ for all } 0\leq a\leq 5;\\[15pt]
\begin{array}{|c||c|}
\hline
a,b & a \\ \hline
a & a \\ 
\hline
\end{array} &
\begin{array}{|c||c|}
\hline
a,b & a,b \\ \hline
a,b & b \\ 
\hline
\end{array} &&\text{ for all } 0\leq a<b\leq 5;\\[15pt]
 \begin{array}{|c||c|}
\hline
a,b,c & a \\ \hline
a & a \\ 
\hline
\end{array} & 
\begin{array}{|c||c|}
\hline
a,b,c & a,b \\ \hline
a,b & a \\ 
\hline
\end{array} &
\begin{array}{|c||c|}
\hline
a,b,c & a,b \\ \hline
a,b & b \\ 
\hline
\end{array} &\\[15pt]
\begin{array}{|c||c|}
\hline
a,b,c & a,b,c \\ \hline
a,b,c & b \\ 
\hline
\end{array} &
\begin{array}{|c||c|}
\hline
a,b,c & a,b,c \\ \hline
a,b,c & c \\ 
\hline
\end{array} & 
 & \text{ for all } 0\leq a<b<c\leq 5.
\end{array}\]
\vspace{0.4cm}
\caption{The 136 exceptional boundary vertices of $\chi^2(\da^5)$.}
\label{table:exv2}
\end{table}
Clearly, there are 136 exceptional vertices. Note, that
$|\supp\sigma|\leq 3$ whenever $v$ is an exceptional boundary vertex.

The general rule for the labeling $I$ is now as follows: 
\[I(v):=\begin{cases}
1, & \text{ if } v \text{ is an internal vertex}; \\
1, & \text{ if } v \text{ is an exceptional boundary vertex};\\
0, & \text{ if } v \text{ is a~regular boundary vertex}.
\end{cases}
\]
Clearly, this labeling is compliant.

We remark that since any $5$-simplex contains an internal vertex,
there are no $0$-mono\-chromatic $5$-simplices in the labeling $I$. Of
course there are quite many $1$-mono\-chromatic $5$-simplices. We will
eliminate them by passing to the third chromatic subdivision.

\subsection{The Matching Lemma}
\vs Our next goal is to produce a perfect matching on the set of
$1$-monochromatic $5$-simplices, so that each pair of matched
$5$-simplices shares a~$4$-simplex. We will start with producing
a~partial matching.

As a warm-up, let us as above consider the graph $\gr_n$, whose set of
vertices is given by $n$-simplices of $\chi(\da^n)$. For $n\geq 1$,
let $f_n$ denote the number of vertices of $\Gamma_n$. We have
$f_1=3$, $f_2=13$ etc. There is an easy recursion
\begin{equation}\label{eq:fnr}
f_n=\binom{n+1}{1}f_{n-1}+\binom{n+1}{2}f_{n-2}+\dots+
\binom{n+1}{n-1}f_1+\binom{n+1}{n}+\binom{n+1}{n+1},
\end{equation} 
see \cite[subsection 4.4]{k1} for a more general formula. One can see
by induction that $f_n$ is always odd. To do this, simply evaluate the
right hand side of \eqref{eq:fnr} modulo 2 and see that
\[f_n\equiv\binom{n+1}{1}+\binom{n+1}{2}+\dots+
\binom{n+1}{n-1}+\binom{n+1}{n}+\binom{n+1}{n+1}\equiv 2^{n+1}-1\equiv 1 \mod 2.\] 
Since $f_n$ is odd we cannot hope for a~perfect
matching. However, it is easy to produce a~number of near-perfect
matchings, each one matching all but one simplex. We need the
following piece of terminology.

\begin{df}\label{df:level}
Let $\Sigma=(x_0,\dots,x_k)$ be a~fixed order on a~nonempty subset of
$[n]$, and let $\sigma=(A_1\s\dots\s A_t)$ be an $n$-simplex of
$\chi(\da^n)$. We shall define the number $\lev(\sigma,\Sigma)\in
[n]$, called the {\bf level of $\sigma$ with respect to $\Sigma$} as
follows. If $\sigma=(A_1\s\dots\s A_{t-k-1}\s x_0\s\dots\s x_k)$, then
$\lev(\sigma,\Sigma)$ is not defined; else $\lev(\sigma,\Sigma):=x_m$,
where $m$ is the highest number $0\leq m\leq k$, such that
$A_{t-k+m}\neq x_m$.
\end{df}

Definition~\ref{df:level} can be rephrased as follows. If the last set
in the ordered set partition $\sigma$ is not $x_k$, the
$\lev(\sigma,\Sigma)=x_k$, else we proceed with the next set in
$\sigma$. If the last set in $\sigma$ is $x_k$, but the penultimate
one is not $x_{k-1}$, then $\lev(\sigma,\Sigma)=x_{k-1}$, else we proceed
with the next set in~$\sigma$. One then repeats this step, moving down
in the ordered set partition $\sigma$. The level is then not defined,
if we went through whole $\Sigma$ and did not stop. This will happen
precisely when $\sigma$ has the form $(A_1\s\dots\s A_{t-k-1}\s
x_0\s\dots\s x_k)$.

We are now ready to define a~special kind of matching, which we shall
call the {\it standard matching}.

\begin{df}  \label{df:sm}
Let $\Sigma=(x_0,\dots,x_n)$ be any fixed order on the set $[n]$.
Recall that $\widehat\Sigma$ is an ordered set partition of $[n]$
associated to $\Sigma$. The {\bf standard matching} $M_\Sigma$
associated to $\Sigma$ is a~bijection
\[M_{\Sigma}:\gr_n\sm\widehat\Sigma\longrightarrow \gr_n\sm\widehat\Sigma,\]
defined by \[\sigma\mapsto\flip(\sigma,\lev(\sigma,\Sigma)).\]
\end{df}

There is a~simple verbal formulation for the standard matching. In
order to match a~simplex $\sigma$, simply flip with respect to the
highest possible color, such that the flip preserves the level
function, see also the proof of Proposition~\ref{prop:sm}.

\begin{figure}[hbt]

  %\begin{picture}(0,0)
  %  \special{psfile=#1.pstex}
  %\end{picture}
  \input{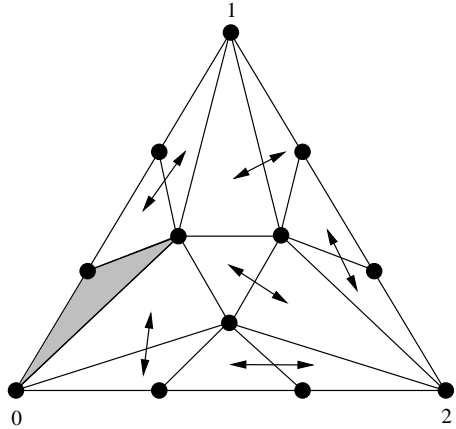}  

\caption{The standard matching on $\mych(\da^2)$ for $\Sigma=(0,1,2)$.}
\label{fig:3}
\end{figure}

The next proposition says that in the standard matching on the set of
all $n$-simplices of $\chi(\da^n)$ with respect to any permutation
$\Sigma$, there is exactly one critical simplex, namely
$\widehat\Sigma$.  This is illustrated by Figure~\ref{fig:3}.

\begin{prop}\label{prop:sm}
The standard matching $M_\Sigma$ described in Definition~\ref{df:sm} is
a~well-defined near-perfect matching on $\gr_n$, with the only critical 
$n$-simplex being indexed by $\widehat\Sigma$.
\end{prop}
\pr Let $\sigma\in \gr_n\sm\widehat\Sigma$. The crucial fact which we need
is that 
\begin{equation}\label{eq:cf3}
\lev(\sigma,\Sigma)=\lev(M_\Sigma(\sigma),\Sigma).
\end{equation}
To see~\eqref{eq:cf3}, assume $\sigma=(A_1\s\dots\s A_m\s
x_{k+1}\s\dots x_n)$, such that $A_m\neq x_k$, i.e.,
$\lev(\sigma,\Sigma)=k$. The ordered set partition $M_\Sigma(\sigma)$
is obtained from $\sigma$ by either merging two sets, or by splitting
one of the sets. If in the process the sets $A_m$, $ x_{k+1}$,
$\dots$, $ x_n$ are untouched, then obviously
$\lev(M_\Sigma(\sigma),\Sigma)=k$ as well. On the other hand, the only
way the sets $ x_{k+1}$, $\dots$, $ x_n$ could be modified would be by
merging them with a~neighbor.  However, this is impossible, since
merges must always involve the set $ x_k$, and we have assumed that
$A_m\neq x_k$.

It remains to consider the case that the set $A_m$ was altered. If it was split 
into $ x_k$ and $A_m\sm x_k$, then again~\eqref{eq:cf3} is satisfied,
since $A_m\sm x_k\neq x_k$. Finally, if the set $A_m$ was merged with
the set $A_{m-1}$, then again $A_{m-1}\cup A_m\neq x_k$, and \eqref{eq:cf3}
is valid again.

Now it follows from \eqref{eq:cf3} together with \eqref{eq:ffx} that
\[M_\Sigma(M_\Sigma(\sigma))=\cf(\cf(\sigma,\lev(\sigma,\Sigma)),\lev(\sigma,\Sigma))
=\sigma.\] This means that we have a well-defined matching, and that
all simplices $\sigma$ for which $\lev(\sigma,\Sigma)$ is defined are
matched. We conclude that we have a~near-perfect matching with the
critical $n$-simplex being indexed by~$\widehat\Sigma$.  \myqed

Note, that even if we have $\Sigma=(x_0,\dots,x_m)$, for $m<n$, the
matching $M_\Sigma$ is still defined for those simplices $\sigma$, for
which the number $\lev(\sigma,\Sigma)$ is defined. In this case, we
will have a~number of critical simplices, namely all those indexed by
the tuples $(A_1\s\dots\s A_k\s x_0\s\dots\s x_m)$, and the proof is
simply the same as that of Proposition~\ref{prop:sm}.

\begin{df}
For a~proper nonempty subset $V\subset[n]$ we let $\gr_n(V)$ denote
the subgraph of $\gr_n$ induced by set of all $n$-simplices
$\sigma=(A_1\s\dots\s A_t)$, such that $A_1\not\subseteq V$.
\end{df}

\begin{prop}\label{prop:sm2}
Let $\Sigma=(x_0,\dots,x_k)$ be the set $[n]\sm V$ taken in an
arbitrary order.  Then $M_\Sigma$ provides a~perfect matching of the
simplices in $\gr_n(V)$.
\end{prop}
\pr The part of the proof of Proposition~\ref{prop:sm} showing \eqref{eq:cf3}
did not use that fact that $\Sigma$ has cardinality $n+1$, so we still have
that identity. This means that $M_\Sigma$ is still a~well-defined matching on 
$\gr_n$, only now with many more critical simplices. 

First, we note that $\lev(\sigma,\Sigma)$ is defined for all $\sigma\in \gr_n(V)$.
Indeed, of $\lev(\sigma,\Sigma)$ is not defined, then $\sigma=(A_1\s\dots\s 
A_k\s x_0\s\dots\s x_m)$, but then $A_1\subseteq A_1\cup\dots\cup A_k=
[n]\sm\{x_0,\dots,x_m\}=V$ yields a~contradiction. Furthermore, we have the implication
\[\sigma\in \gr_n(V)\Rightarrow M_\Sigma(\sigma)\in \gr_n(V).\]
Indeed, assume $M_\Sigma(\sigma)=(B_1\s\dots\s B_q)$, and
$\sigma=(A_1\s\dots\s A_t)$, with $A_1\not\subseteq V$. Then, by
definition of $M_\Sigma$, either we have $B_1\supseteq A_1$, or $B_1=
x_k$, where $x_k=\lev(\sigma,\Sigma)\in[n]\sm V$.  Either way, we get
$B_1\not\subseteq V$, so $M_\Sigma(\sigma)\in \gr_n(V)$.  We now
conclude that $M_\Sigma$ is a~perfect matching on $\gr_n(V)$.  \myqed

\begin{df}\label{df:st} 
Given an $n$-simplex $\sigma=(A_1\s\dots\s A_t)$, with
$A_t=\{x_1,\dots,x_k\}$, such that $x_1<\dots< x_k$, we set
$\st(\sigma):=(x_1,\dots,x_k)$.
\end{df}

\begin{df}
Given a subset $V\subset[n]$, we shall call a~tuple
$\ca=(A_1\s\dots\s A_k)$ of nonempty subsets of $V$ a~{\bf prefix} in~$V$, 
if $A_1,\dots,A_k$ are disjoint. We shall say that a prefix is
{\bf full} if $A_1\cup\dots\cup A_k=V$. We also allow an empty prefix.

Furthermore, given a prefix $\ca=(A_1\s\dots\s A_k)$, we let $\gr_n(V,\ca)$
denote the subgraph of $\gr_n$ induced by the set of all $n$-simplices
$\sigma=(A_1\s\dots\s A_k\s A_{k+1}\s\dots\s A_t)$ of $\chi(\da^n)$, such that
$A_{k+1}\not\subseteq V$.
\end{df}

\begin{figure}[hbt]

  %\begin{picture}(0,0)
  %  \special{psfile=#1.pstex}
  %\end{picture}
  \input{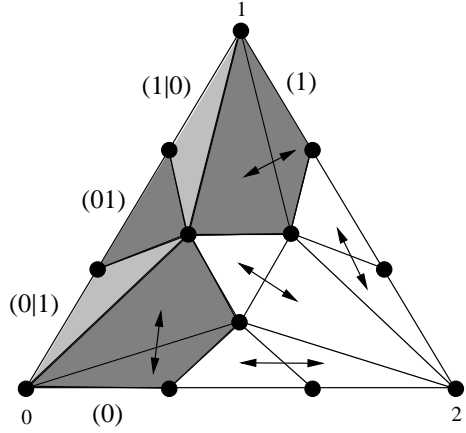}  

\caption{The decomposition of $\gr_2$ for $V=\{0,1\}$.}
\label{fig:4}
\end{figure}

In particular, the set $\gr_n(V,\emptyset)$ coincides with the
previously defined set $\gr_n(V)$. We note, that for any
$V\subset[n]$, the set of vertices of $\gr_n$, which is the set of all
$n$-simplices of $\chi(\da^n)$, is a~disjoint union of the sets of
vertices of $\gr_n(V,\ca)$, where $\ca$ runs over all prefixes in~$V$.
The following lemma, whose statement is illustrated by
Figure~\ref{fig:4}, provides a useful generalization of
Proposition~\ref{prop:sm2}.

\begin{lm}\label{lm:sml}
Assume $V\subset[n]$, and $\Sigma=(x_0,\dots,x_m)$ is an~arbitrary
permutation of the set $[n]\sm V$. Let furthermore
$\ca=(A_1\s\dots\s A_k)$ be an~arbitrary prefix of~$V$.
\begin{enumerate}
\item[(1)] If the prefix $\ca$ is not full, then the standard matching
$M_\Sigma$ gives a~perfect matching of simplices in $\gr_n(V,\ca)$.
\item[(2)] If the prefix $\ca$ is full then the standard matching
  $M_\Sigma$ of $\gr_n(V,\ca)$ has precisely one critical $n$-simplex,
  namely $\sigma=(A_1\s\dots\s A_k\s x_0\s\dots\s x_m)$.
\end{enumerate}
\end{lm}
\pr Our argument is very close to the proof of
Proposition~\ref{prop:sm2}.  Take $\sigma\in \gr_n(V,\ca)$. If
$\lev(\sigma,\Sigma)$ is not defined, then $\sigma$ has the form
$(B_1\s\dots\s B_t\s x_0\s\dots\s x_m)$. Since $\sigma\in
\gr_n(V,\ca)$, we see that $\sigma=(A_1\s\dots\s A_k\s x_0\s\dots\s
x_m)$, which is only possible if the prefix $\ca$ is full. So, if
$\ca$ is full, we do have one critical simplex, otherwise we do not.

Assume now $\sigma=(A_1\s\dots\s A_k\s A_{k+1}\s\dots\s A_t)$, such
that $\sigma\neq(A_1\s\dots\s A_k\s x_0\s\dots\s x_m)$, and let
$x_t=\lev(\sigma,\Sigma)$. We have $A_{k+1}\cap V\neq\emptyset$.  The
ordered set partition $M_\Sigma(\sigma)$ is obtained from $\sigma$ by
merging two sets or splitting a~set. If the sets $A_1,\dots,A_{k+1}$
are untouched, then clearly $M_\Sigma(\sigma)\in\Gamma_n(V,\ca)$.  In
the other hand, the sets $A_1,\dots,A_k$ cannot be involved in any
modification, so the only remaining case is that $A_{k+1}$ has been
modified. Either $A_{k+1}$ has been replaced by $x_t$ followed by
$A_{k+1}\sm x_t$, or $A_{k+1}=x_t$ and we replace $A_{k+1}$, $A_{k+2}$
with $x_t\cup A_{k+2}$. In any case, $x_t\in B_{k+1}$, where
$M_\Sigma(\sigma)=(A_1\s\dots\s A_k\s B_{k+1}\s\dots)$, hence $V\cup
B_{k+1}\neq\emptyset$. Thus, we derive the implication
\[\sigma\in \gr_n(V,\ca)\Rightarrow M_\Sigma(\sigma)\in \gr_n(V,\ca),\]
for non-critical $\sigma$. Hence $M_\Sigma$ gives a~perfect matching on the
set of all $n$-simplices in $\gr_n(V,\ca)$, with the only possible
exception being  $\sigma=(A_1\s\dots\s A_k\s x_0\s\dots\s x_m)$.
\qed

\subsection{Using the Matching Lemma to get a partial matching}
\label{ss:3.4}
\vs
Before we proceed with describing the partial matching on the set of
all $1$-monochromatic $5$-simplices of the labeling $I$, let us give
combinatorial description of all the elements in this set. To start
with, all $5$-simplices with only internal vertices are
$1$-monochromatic. These are described by a simple criterion in
Proposition~\ref{prop:int}. The rest of $1$-monochromatic
$5$-simplices are those which contain internal vertices and some
exceptional boundary vertices.  In the Table~\ref{table:exc-vert} we
list all exceptional boundary vertices and the $5$-simplices which
contain them.

\begin{table}[hbt]
\[
\begin{array}{l|l}
\text{exceptional vertices }& $5$-\text{simplices with modified labels}  
\\ \hline 
\\%[15pt] 
\begin{array}{|c||c|}
\hline
a & a \\ \hline
a & a \\ 
\hline
\end{array} & 
\begin{array}{l}
(a\s S_1\s\dots\s S_p\ds a\s T_1\s \dots\s T_q)
\end{array}
\\ \\
\begin{array}{|c||c|}
\hline
a,b & a \\ \hline
a & a \\ 
\hline
\end{array} & 
\begin{array}{l}
(a,b\s S_1\s \dots\s S_p\ds a\s T_1\s \dots\s T_q)\\
(b\s a\s S_1\s \dots\s S_p\ds a\s T_1\s \dots\s T_q)
\end{array}
\\ \\
\begin{array}{|c||c|}
\hline
a,b & a,b \\ \hline
a,b & b \\ 
\hline
\end{array} & 
\begin{array}{l}
(a,b\s S_1\s \dots\s S_p\ds a,b\s T_1\s \dots\s T_q)\\
(a,b\s S_1\s \dots\s S_p\ds a\s b\s T_1\s \dots\s T_q)
\end{array}
\\
\\
\begin{array}{|c||c|}
\hline
a,b,c & a \\ \hline
a & a \\ 
\hline
\end{array} & 
\begin{array}{l}
(a,b,c\s S_1\s \dots\s S_p\ds a\s T_1\s \dots\s T_q)\\
(c\s a,b\s S_1\s \dots\s S_p\ds a\s T_1\s \dots\s T_q)\\
(b\s a,c\s S_1\s \dots\s S_p\ds a\s T_1\s \dots\s T_q)\\
(b,c\s a\s S_1\s \dots\s S_p\ds a\s T_1\s \dots\s T_q)\\
(b\s c\s a\s S_1\s \dots\s S_p\ds a\s T_1\s \dots\s T_q)\\
(c\s b\s a\s S_1\s \dots\s S_p\ds a\s T_1\s \dots\s T_q)
\end{array}
\\ \\
\begin{array}{|c||c|}
\hline
a,b,c & a,b \\ \hline
a,b & a \\
\hline
\end{array} & 
\begin{array}{l}
(a,b,c\s S_1\s   \dots\s S_p\ds a,b\s   T_1\s \dots\s T_q)\\
(c\s a,b\s S_1\s \dots\s S_p\ds a,b\s   T_1\s \dots\s T_q)\\
(a,b,c\s S_1\s   \dots\s S_p\ds b\s a\s T_1\s \dots\s T_q)\\
(c\s a,b\s S_1\s \dots\s S_p\ds b\s a\s T_1\s \dots\s T_q)
\end{array}
\\ \\
\begin{array}{|c||c|}
\hline
a,b,c & a,b \\ \hline
a,b & b \\
\hline
\end{array} & 
\begin{array}{l}
(a,b,c\s S_1\s \dots\s S_p  \ds a,b\s   T_1\s \dots\s T_q)\\
(c\s a,b\s S_1\s \dots\s S_p\ds a,b\s   T_1\s \dots\s T_q)\\
(a,b,c\s S_1\s \dots\s S_p  \ds a\s b\s T_1\s \dots\s T_q)\\
(c\s a,b\s S_1\s \dots\s S_p\ds a\s b\s T_1\s \dots\s T_q)
\end{array}\\ \\
\begin{array}{|c||c|}
\hline
a,b,c & a,b,c \\ \hline
a,b,c & c \\ 
\hline
\end{array} & 
\begin{array}{l}
(a,b,c\s S_1\s \dots\s S_p\ds a,b,c\s T_1\s \dots\s T_q)\\
(a,b,c\s S_1\s \dots\s S_p\ds a,b\s c\s T_1\s \dots\s T_q)\\
(a,b,c\s S_1\s \dots\s S_p\ds b\s a\s c\s T_1\s \dots\s T_q)\\
(a,b,c\s S_1\s \dots\s S_p\ds a\s b\s c\s T_1\s \dots\s T_q)\\
(a,b,c\s S_1\s \dots\s S_p\ds b\s a,c\s T_1\s \dots\s T_q)\\
(a,b,c\s S_1\s \dots\s S_p\ds a\s b,c\s T_1\s \dots\s T_q)
\end{array} \\ \\
\begin{array}{|c||c|}
\hline
a,b,c & a,b,c \\ \hline
a,b,c & b \\ 
\hline
\end{array} & 
\begin{array}{l}
(a,b,c\s S_1\s \dots\s S_p\ds a,b,c\s T_1\s \dots\s T_q)\\
(a,b,c\s S_1\s \dots\s S_p\ds a,c\s b\s T_1\s \dots\s T_q)\\
(a,b,c\s S_1\s \dots\s S_p\ds c\s a\s b\s T_1\s \dots\s T_q)\\
(a,b,c\s S_1\s \dots\s S_p\ds a\s c\s b\s T_1\s \dots\s T_q)\\
(a,b,c\s S_1\s \dots\s S_p\ds c\s a,b\s T_1\s \dots\s T_q)\\
(a,b,c\s S_1\s \dots\s S_p\ds a\s b,c\s T_1\s \dots\s T_q)
\end{array}
\end{array}\]

\vspace{0.4cm}
\caption{Exceptional boundary vertices and $5$-simplices containing them;
here we assume that $a<b<c$.}
\label{table:exc-vert}
\end{table}

%Here was table:exc-vert

Next we need to understand which of the $5$-simplices from the right
column of the Table~\ref{table:exc-vert} contain {\it only}
exceptional boundary vertices.  Recall that a~$5$-simplex of $\kom$
is indexed by a~pair of ordered set partitions of subsets of $[5]$.
We shall describe the set of $5$-simplices containing only exceptional
boundary vertices by grouping them together according to the first
ordered set partition. 

\begin{df}
For a~$5$-simplex $\sigma$ of $\chi(\da^5)$, let $\cn(\sigma)$ denote
the subgraph of $\Gamma_n$ induced by all $5$-simplices $\tau$ of
$\chi(\da^5)$ such that the $5$-simplex $(\sigma\ds\tau)$ of $\kom$ is
$1$-monochromatic.
\end{df}

\begin{table}[hbt]
\[\begin{array}{l|l}
\text{type of }\sigma& \text{allowed prefixes} \\ \hline \\
(a\s S_1\s\dots\s S_p) & (a) \\ \\
(ab\s S_1\s\dots\s S_p) & (a),\quad (a\s b) \\ \\
(abc\s S_1\s\dots\s S_p) & (a),\quad (a\s b),\quad (ab),\quad
(ab\s c),\quad (a\s bc),\quad (a\s b\s c) \\ \\
(a\s b\s S_1\s\dots\s S_p) & (a) \\ \\
(b\s a\s S_1\s\dots\s S_p) &  (a),\quad (b) \\ \\
(ab\s c\s S_1\s\dots\s S_p) & (a),\quad (a\s b)\\ \\
(ac\s b\s S_1\s\dots\s S_p) & (a),\quad (a\s c) \\ \\
(bc\s a\s S_1\s\dots\s S_p) & (a),\quad (b),\quad (b\s c) \\ \\
(a\s bc\s S_1\s\dots\s S_p) & (a) \\ \\
(b\s ac\s S_1\s\dots\s S_p) & (a),\quad (b) \\ \\
(c\s ab\s S_1\s\dots\s S_p) & (a),\quad (c),\quad (ab),\quad (a\s b) \\ \\
(a\s b\s c\s S_1\s\dots\s S_p) & (a) \\ \\
(a\s c\s b\s S_1\s\dots\s S_p) & (a) \\ \\
(b\s a\s c\s S_1\s\dots\s S_p) & (a),\quad (b) \\ \\
(b\s c\s a\s S_1\s\dots\s S_p) & (a),\quad (b) \\ \\
(c\s a\s b\s S_1\s\dots\s S_p) & (a),\quad (c) \\ \\
(c\s b\s a\s S_1\s\dots\s S_p) & (a),\quad (b), \quad (c) 
\end{array}\]
\vspace{0.4cm}
\caption{New $1$-monochromatic $5$-simplices sorted by the supporting
  $5$-simplex in $\chi(\da^5)$; here we assume $a<b<c$.}
\label{table:exc-simp}
\end{table}

The explicit description of the graph $\cn(\sigma)$ is given in 
Proposition~\ref{prop:3.11}, using Table~\ref{table:exc-simp}.
We remark that the types of the $5$-simplices listed in the left 
column of that table are meant to be mutually exclusive.
%Specifically, it means that there is an additional condition for the
%$5$-simplex in the first row: if $|S_1|\leq 2$, then $a<\max S_1$.

\begin{prop} \label{prop:3.11}
Let $\sigma=(S_1\s \dots\s S_p)$ be a~$5$-simplex of $\chi(\da^5)$,
and set $V:=[5]\sm S_p$. Then we have
$\cn(\sigma)=\gr_5(V)\cup\Lambda$, where the subset $\Lambda$ is given
as follows. If $\sigma$ is of the type listed on the left column of
Table~\ref{table:exc-simp}, then
\begin{equation}\label{eq:lam}
\Lambda=\bigcup_\ca \gr_5(V,\ca),
\end{equation}
where the union is taken over all prefixes $\ca$ which are listed in 
the corresponding row of Table~\ref{table:exc-simp}.
We set $\Lambda$ to be an empty set otherwise.
\end{prop}
\pr To start with, we see that $V\subset[n]$. If $V$ is empty, then 
we are looking at the central $5$-simplex $\sigma=([5])$, in which case 
all of the $5$-simplices of $\cn(\sigma)$ are internal. 

Assume now $V$ is non-empty. Let $\tau=(T_1\s\dots\s T_q)$ be 
a~$5$-simplex of $\chi(\da^5)$. By Proposition~\ref{prop:int}, 
all the vertices of the $5$-simplex $(\sigma\ds\tau)$ of $\chi^2(\da^5)$ 
are internal if and only if $S_p\cap T_1\neq\emptyset$, or 
equivalently $T_1\not\subseteq V$. In other words, the 5-simplex
$(\sigma\ds\tau)$ is internal if and only if $\tau\in \gr_5(V)$.
All internal $5$-simplices are 1-monochromatic, and all further 
$1$-monochromatic $5$-simplices appear due to the exceptional vertices.
So if did not have any, we would simply have $\cn(\sigma)=\gr_5(V)$.

%\newpage

% Here was table exc-simp

To take into account the influence of the exceptional vertices, we 
obtain the correction set $\Lambda$ by a~direct case-by-case analysis.  
Table~\ref{table:exc-vert} shows the cases which need to be
considered. The Table~\ref{table:exc-simp} describes in each one of
these cases, which new $1$-monochromatic simplices will arise.
\qed

\section{Modifying the standard matching}\label{sect:4}
\subsection{Conducting graphs}
\vs In order to look at matchings in depth, we need certain standard
tools from the matching theory. Given a partial matching, there is a
classical way of modifying it, possibly enlarging, or even making it
perfect. Following definition provides the key concepts.
\begin{df}
Given a matching $M$ on a~graph $G$, assume $\gamma$ is an
edge path in $G$ with endpoints $v$ and $w$. We call the path $\gamma$
\begin{itemize}
\item {\bf alternating} if the edges of $\gamma$ are alternatively
  matching and non-matching;
\item {\bf properly alternating} if is alternating and those endpoints of
the path which are not matched on the path are critical; 
\item {\bf augmenting} if it is properly alternating, starting and ending with 
non-matching edges;
\item {\bf semi-augmenting} if it is properly alternating, starting
  with a~matching edge and ending with a non-matching one;
\item {\bf weakly semi-augmenting} if it is alternating, starting
  with a~matching edge and ending with a non-matching one;
\item {\bf non-augmenting} if it is alternating, starting and
  ending with matching edges.
\end{itemize}
\end{df}

Properly alternating non-self-intersecting edge paths allow us to
modify matchings.

\begin{df}\label{df:mdef}
Assume we are given a matching $M$ on a~graph $G$, and a properly
alternating non-self-intersecting edge path $\gamma$. We define
$D(M,\gamma)$ be a new matching on $G$ consisting of all edges from
$M$ which do not belong to $\gamma$ together with all edges from
$\gamma$ which do not belong to $M$.
\end{df}

The next proposition shows that having a~properly alternating edge
path is enough to find a larger matching, and eventually to turn
partial matchings into perfect ones.

\begin{prop}\label{prop:ap}
Assume $M$ is a matching on a~bipartite graph~$G$, and $v,w$ are
critical vertices of~$G$ with respect to~$M$. Assume furthermore, that
there exists a~properly alternating edge path $\gamma$ from $v$
to~$w$. Then, there exists a~matching $\wti M$ on $G$, such that the
set of critical vertices with respect to $\wti M$ is obtained from the
set of critical vertices with respect to $M$ by removing the vertices
$v$ and~$w$.
\end{prop}
\pr If the path $\gamma$ is non-self-intersecting, then the new
matching $\wti M$ can be taken to be $D(M,\gamma)$. Assume $\gamma$ is
a~self-intersecting path, and its vertices, listed in the path order,
are $v=v_0$, $v_1$, $\dots$, $v_l=w$. 

Since $\gamma$ is self-intersecting, there exist $0\leq k<m\leq l$,
such that $v_k=v_m$, and there is no vertex duplication in the
sequence $v_k$, $v_{k+1}$, $\dots$, $v_{m-1}$. Let $\tilde\gamma$ be
obtained from $\gamma$ by removing the loop consisting of the edges
$(v_k,v_{k+1})$, $(v_{k+1},v_{k+2})$, $\dots$, $v_{m-1},v_m$. Since
the graph $G$ is bipartite, that loop must have an~even length, hence
the new path $\tilde\gamma$ is again properly alternating. Repeating
this procedure we will eventually arrive at a~properly alternating
non-self-intersecting edge path connnecting $v$ with~$w$.  
\qed

%It is easy to see that $D(M,\gamma)$ is a matching again. If $\gamma$
%is augmenting, the number of critical vertices decreases by $2$,
%making this procedure useful when trying to turn partial matchings
%into perfect ones. 

We note, that if the edge path $\gamma$ in Proposition~\ref{prop:ap}
is semi-augmenting, then the number of critical vertices does not
change, whereas, if $\gamma$ is non-augmenting, then the number of
critical vertices increases by~$2$.  Even though modifying a~matching
along a~semi-augmenting path does not change the number of critical
vertices, this modification is still useful as it could be thought of
as transporting the critical vertex from one endpoint of the
semi-augmenting path to the other.

\begin{df} \label{df:strm}
Let $G$ be a~bipartite graph, with the corresponding bipartite
decomposition $(A,B)$. We say that $G$ is {\bf conducting} if one of the 
following situations occur: 
\begin{enumerate}
\item[(1)] We have $|A|=|B|+1$, and for any vertex $v\in A$, the graph $G$
has a~near-perfect matching, with the critical vertex~$v$.
\item[(2)] We have $|A|=|B|$, and for any vertices $v\in A$, $w\in B$,
  the graph $G$ has a~matching, with precisely two critical
  vertices~$v$ and~$w$.
\end{enumerate}
More specifically, if a~conducting graph $G$ satisfies condition (1)
we shall call it {\bf conducting graph of the first type}, else
we  shall call it {\bf conducting graph of the second type}.
\end{df}

One handy way of showing that a~graph is conducting is by presenting
a certain collection of paths, as the next proposition details.

\begin{prop}\label{prop:bg}
Let $G$ be a~bipartite graph, with the corresponding bipartite
decomposition $(A,B)$.
\begin{enumerate}
\item[(1)] Assume $M$ is a~near-perfect matching of $G$ with critical
  vertex $v\in A$, such that for any other vertex $w\in A$ there
  exists a~semi-augmenting path from $v$ to~$w$, then the graph $G$ is
  conducting of the first type.
\item[(2)] Assume $M$ is a~perfect matching of $G$, such that for any
  vertices $v\in A$, $w\in B$ there exists a~non-augmenting path from
  $v$ to~$w$, then the graph $G$ is conducting of the second type.
\end{enumerate}
\end{prop}
\pr This is an immediate consequence of Proposition~\ref{prop:ap}.
\myqed

\begin{df}\label{df:closed}
Let $\Omega$ be a~family of prefixes of a set $V\subset[n]$ containing
the empty prefix. We say that $\Omega$ is {\bf closed} if the following 
two conditions are satisfied:
\begin{enumerate}
\item[(1)] if $(A_1\s\dots\s A_k\s x)\in\Omega$, where
  $A_1,\dots,A_k\subseteq V$, $x\in V$, then $(A_1\s\dots\s A_k)\in\Omega$;
\item[(2)] if $(A_1\s\dots\s A_k)\in\Omega$, and $|A_m|\geq 2$, for some 
$1\leq m\leq k$, then there exists $x\in A_m$, such that 
\[(A_1\s\dots\s A_{m-1}\s x\s A_m\sm x\s A_{m+1}\s\dots\s A_k)\in\Omega.\]
\end{enumerate}
Furthermore, we let $\gr_n(V,\Omega)$ denote the subgraph of
$\Gamma_n$ induced by the vertices of
$\bigcup_{\ca\in\Omega}\gr_n(V,\ca)$.
\end{df}

\begin{rem}\label{rem:sn}
By Lemma~\ref{lm:sml}, the standard matching will induce a~matching on
$\gr_n(V,\Omega)$. However, this matching does not have to be perfect
or near-perfect. As a~matter of fact, the number of critical vertices
will be equal to the number of full prefixes in~$\Omega$.
\end{rem}

The next observation can be shown by direct inspection.

\begin{rem}\label{rem:cl} 
All families of prefixes $\Omega$ listed in Table~\ref{table:exc-simp}
are closed.
\end{rem}

We shall now proceed with analysing conductibility of graphs
$\gr_n(V,\Omega)$. Let us fix a~natural number~$n\geq 3$. Let $\rho$ denote
the central simplex of $\chi(\da^n)$, indexed by $([n])$, and let
$\nu$ denote the neighboring $n$-simplex indexed by $(
n,[n-1])$. Consider the standard matching $M_\Sigma$, associated to
$\Sigma=(0,1,\dots,n)$. The only critical simplex of $M_\sigma$ is
$(0\s 1\s\dots\s n)$, and clearly, $M_\sigma$ matches $\rho$
with~$\nu$.  Our central tool will be the following techical lemma.

\begin{lm}\label{lm:1}
Let $\sigma$ be a~non-critical $n$-simplex in $\chi(\da^n)$, for $n\geq 3$. There
exists a weakly semi-augmenting (with respect to $M_\Sigma$) path $P$
starting at $\sigma$ and ending either at $\rho$ or
at~$\nu$. 
%Furthermore, the path $P$ can be chosen to have no self-intersections.
\end{lm}

\pr Since the endpoints of a~weakly semi-augmenting path have the same
orientation, we know that if such a path $P$ exists, then its endpoint
from the set $\{\rho,\nu\}$ is uniquely determined. For brevity, we
shall say that $P$ {\it goes from $\sigma$ to $\{\rho,\nu\}$.} Taking the
empty path we see that the claim is true for $\sigma=\rho$ and for
$\sigma=\nu$. Assume that $\sigma\neq\rho$, $\sigma\neq\nu$. We shall
use the notation $\sigma=(A_1\s\dots\s A_t)$ throughout the proof.

When describing edges in a~path, we shall usually not
list entire combinatorial indices of the simplices, but rather the
part of the ordered set partition where the actual change occurs.
The path from $\sigma$ to $\{\rho,\nu\}$ will be constructed by
concatenating different pieces. 

Let us briefly introduce some notations in which we encode our paths.
Assume $\sigma=(A_1\s\dots\s A_t)$ is an $n$-simplex of $\chi(\da^n)$,
and set $k:=\lev(\sigma,\Sigma)$. 
\begin{itemize}
\item We let $\stackrel{k}{\mra}$ denote the matching edge between $\sigma$ and
  $M_\Sigma(\sigma)=\cf(\sigma,k)$. 
\item Assume $A_i=\{x\}$, for some $1\leq i\leq t-1$, and $x\neq k$.
  We let $\stackrel{x\cup}{\sra}$ denote the edge between $\sigma$
  and $(A_1\s\dots\s A_{i_1}\s x\cup A_{i+1}\s\dots\s A_t)$.
\item Assume instead that $|A_i|\geq 2$, for some $1\leq i\leq t$,
  $x\in A_i$, and $x\neq k$. We let $\stackrel{x\s}{\sra}$ denote
  the edge between $\sigma$ and $(A_1\s\dots\s A_{i_1}\s x\s
  A_i\sm x\s\dots\s A_t)$.
\item If $|A_i|\geq 2$, for some $1\leq i\leq t$, and
  $\lev(\sigma,\Sigma)\notin A_i$, we use the
  notation~$\stackrel{A_i}{\sra}$ to denote the set of edges
  $\stackrel{x\s}{\sra}$, for all $x\in A_i$.
\end{itemize}
We shall call the move $\stackrel{A}{\sra}$ the {\it generic split}.
When a generic split is used in our notation of the path, it means
that it does not matter at this point which of the elements of $A$ we
split off, and any will do. Formally, this means that we output a set
of paths, rather than one path. Later, we will need to choose paths
satisfying certain additional conditions. This can be done from this
set of paths, by substituting specific splits instead of generic
ones. For example, condition (2) of Definition $\ref{df:closed}$ only
guarantess the existance of {\it some} element $x$, which we are
allowed to split. If at this point we used generic split of $A_m$ in
our path, then we will be able to pick a suitable path from our set.

%When $n=1$ or $n=2$, the claim can be proved by direct inspection. 
We break our argument into considering three different cases.

\vskip5pt

\nin {\bf Case 1.} Assume that $\lev(\sigma,\Sigma)=n$. Pick $i$ such that
$n\in A_i$. Our assumption means that either $A_i= n$, and $i\leq t-1$, 
or $|A_i|\geq 2$. 

Our first goal, is to reduce this to the case $i=1$. Assume that $i\geq 2$,
and $|A_{i-1}|=1$, say $A_{i-1}= a$. We can use the moves
\eqref{eq:m1} and \eqref{eq:m2}  to lower the index $i$ by $1$.
\begin{equation}\label{eq:m1}
(\dots\s a\s  n\cup B\s\dots) \mran 
(\dots\s a\s n\s B\s\dots) \stackrel{a\cup}\sra 
(\dots\s\{n,a\}\s B\s\dots)
\end{equation}
\begin{equation}\label{eq:m2}
(\dots\s a\s n\s B\s\dots) \mran (\dots\s a\s  n\cup B\s\dots) 
\stackrel{a\cup}\sra (\dots\s \{n,a\}\cup B\s\dots) 
\end{equation}

Assume now $i\geq 2$, and $|A_{i-1}|\geq 2$. We can use the moves
\eqref{eq:m3} and \eqref{eq:m4}  to lower the cardinality $|A_{i-1}|$ by $1$,
eventually reducing this to the case above.
\begin{equation}\label{eq:m3}
(\dots\s A_{i-1}\s  n\cup B\s\dots)\mran 
(\dots\s A_{i-1}\s n\s B\s\dots) \stackrel{A_{i-1}}\sra 
(\dots\s x\s A_{i-1}\sm x\s n\s B\s\dots)
\end{equation}
\begin{equation}\label{eq:m4}
(\dots\s A_{i-1}\s n\s B\s\dots)\mran 
(\dots\s A_{i-1}\s  n\cup B\s\dots)\stackrel{A_{i-1}}\sra 
(\dots\s x\s A_{i-1}\sm x\s  n\cup B\s\dots)
\end{equation}
In either case, we are able to lower the index $i$ by $1$, and in the
end to reach the case $n\in A_1$.

Next, we want to achieve $|A_1|=2$, or $|A_1|=|A_2|=1$.
Assume first $|A_1|\geq 3$, then use the move~\eqref{eq:m5} to achieve  
$|A_1|=|A_2|=1$.
\begin{equation}\label{eq:m5}
(n\cup A\s\dots)\mran(n\s A\s\dots)\stackrel{A}\sra 
(n\s x\s A\sm x\s\dots)
\end{equation}

Assume now $|A_1|=1$, but $|A_2|\geq 2$. Since $\sigma\neq\nu$, 
we must have $|A_3|\geq 1$. If $|A_3|\geq 2$, then use the move \eqref{eq:m6}
to reduce to the case $|A_1|\geq 3$, which we just dealt with. 
\begin{equation}\label{eq:m6}
(n\s A_2\s A_3\s\dots)\mran(n\cup A_2\s A_3\s\dots)\stackrel{A_3}\sra 
(n\cup A_2\s x\s A_3\sm x\s\dots)
\end{equation}
If $|A_3|=1$ and $A_4\neq\emptyset$, then use the move \eqref{eq:m7}
to reduce to the case $|A_1|\geq 3$ again.
\begin{equation}\label{eq:m7}
(n\s A_2\s b\s A_4\s\dots)\mran 
(n\cup A_2\s b\s A_4\s\dots)\stackrel{b\cup}\sra 
(n\cup A_2\s b\cup A_4\s\dots)
\end{equation}

Finally, assume $|A_3|=1$, $A_4=\emptyset$. Then, use the move \eqref{eq:m8},
picking $a\in\{n-1,n-2\}\cap A_2$, to reduce to the case $|A_1|=2$.
\begin{multline}\label{eq:m8}
(n\s A_2\s b)\mran 
(n\cup A_2\s b)\stackrel{a\s}\sra 
(a\s (n\cup A_2)\sm a\s b)\\ \mran
(a\s n\s A_2\sm a\s b)\stackrel{a\cup}\sra 
(\{n,a\}\s A_2\sm a\s b)
\end{multline}
So at this point we have $n\in A_1$, and  $|A_1|=2$, or $|A_1|=|A_2|=1$.

Using one of the moves \eqref{eq:m9} and \eqref{eq:m10}, whenever
$|A_k|\geq 2$, we will eventually arrive at the simplex which is
either encoded by $(n\s a_1\s\dots\s a_n)$ or by $(\{n,a_1\}\s
a_2\s\dots\s a_n)$.
\begin{equation}\label{eq:m9}
(n\s a_1\s\dots\s A_k\s\dots)\mran 
(\{n,a_1\}\s\dots\s A_k\s\dots)\stackrel{A_k}\sra 
(\{n,a_1\}\s\dots\s x\s A_k\sm x \s \dots)
\end{equation}
\begin{equation}\label{eq:m10}
(\{n,a_1\}\s\dots\s A_k\s\dots)\mran 
(n\s a_1\s\dots\s A_k\s\dots)\stackrel{A_k}\sra 
(n\s a_1\s\dots\s x\s A_k\sm x \s \dots)
\end{equation}
 
We now continue with moves \eqref{eq:m11} and \eqref{eq:m12}
reducing our simplex either to $(\{n,a\}\s A)$ or to $(n\s a\s A)$.
\begin{equation}\label{eq:m11}
(n\s a_1\s\dots\s a_t\s A)\mran 
(\{n,a_1\}\s a_2\s\dots\s a_t\s A)\stackrel{a_t\cup}\sra 
(\{n,a_1\}\s a_2\s\dots\s a_{t-1}\s \{a_t\}\cup A)
\end{equation}
\begin{equation}\label{eq:m12}
(\{n, a_1\}\s\dots\s a_t\s A)\mran 
(n\s a_1\s a_2\s\dots\s a_t\s A)\stackrel{a_t\cup}\sra 
(n\s a_1\s a_2\s\dots\s a_{t-1}\s \{a_t\}\cup A)
\end{equation} 
If $\sigma=(\{n,a\}\s A)$, we use the move \eqref{eq:m13} to arrive at $\nu$.
\begin{equation}\label{eq:m13}
(\{n,a\}\s A)\mran (n\s a\s A)\stackrel{a\cup}\sra (n\s[n-1])
\end{equation}
So we can assume that $\sigma=(n\s a\s A)$. If $a=n-1$, we use the 
move~\eqref{eq:m14} to arrive at $\rho$.
\begin{equation}\label{eq:m14}
(n\s n-1\s A)\mran (\{n-1,n\}\s A)\stackrel{n-1\s}\sra 
(n-1\s n\s A)\mran (n-1\s  n\cup A) \stackrel{n-1\cup}\sra([n])
\end{equation}
Else write $A= n-1\cup B$ and use  the 
move~\eqref{eq:m15} to arrive at the considered case $a=n-1$.
\begin{multline}\label{eq:m15}
(n\s a\s A)\mran (\{n,a\}\s A)\stackrel{n-1\s}\sra 
(\{n,a\}\s n-1\s B) \mran (n\s a\s n-1\s B)\\ \stackrel{a\cup}\sra 
(n\s \{n-1,a\}\s B) \mran(\{n,n-1,a\}\s B)  \stackrel{n-1\s}\sra 
(n-1\s\{n,a\}\s B)\mran (n-1\s n\s a\s B)\\ \stackrel{n-1\cup}\sra 
(\{n-1,n\}\s a\s B)\mran(n\s n-1\s a\s B) \stackrel{a\cup}\sra 
(n\s n-1\s a\cup B)
\end{multline}

\vskip5pt

\nin {\bf Case 2.} Assume $\lev(\sigma,\Sigma)\leq n-2$.
In this case we have $A_{t-1}= n-1$ and $A_t= n$. We use the move 
\eqref{eq:m16} to reduce it to Case 1, where $k=\lev(\sigma,\Sigma)$.
\begin{equation}\label{eq:m16}
(\beta\s n-1\s n)\stackrel{k}\mra(\tilde\beta\s n-1\s n)
\stackrel{n-1\cup}\sra (\tilde\beta\s \{n-1, n\})
\end{equation}

\vskip5pt

\nin {\bf Case 3.} Assume $\lev(\sigma,\Sigma)=n-1$.
If $\sigma=(\dots\s n-1\cup B\s\dots\s A\s n)$, iteratively use the moves 
\eqref{eq:m17} and \eqref{eq:m18} to reduce it to Case 1.
\begin{multline}\label{eq:m17}
(\dots\s  n-1\cup B\s\dots\s A\s n)\stackrel{n-1}\mra
(\dots\s n-1\s B\s\dots\s A\s n) \\
\stackrel{A}\sra 
(\dots\s n-1\s B\s\dots\s x\s A\sm x\s n),\text{ where }|A|\geq 2.
\end{multline}
\begin{multline}\label{eq:m18}
(\dots\s  n-1\cup B\s\dots\s a\s n)\stackrel{n-1}\mra
(\dots\s n-1\s B\s\dots\s a\s n)\\ \stackrel{a\cup}\sra 
(\dots\s n-1\s B\s\dots\s \{n,a\})
\end{multline}
If $\sigma=(\dots\s n-1\s B\s\dots\s A\s n)$, iteratively use the moves 
\eqref{eq:m19} and \eqref{eq:m20} to reduce it to Case 1.
\begin{multline}\label{eq:m19}
(\dots\s n-1\s B\s\dots\s A\s n)\stackrel{n-1}\mra
(\dots\s  n-1\cup B\s\dots\s A\s n) \\
\stackrel{A}\sra 
(\dots\s  n-1\cup B\s\dots\s x\s A\sm x\s n),\text{ where }|A|\geq 2.
\end{multline}
\begin{multline}\label{eq:m20}
(\dots\s n-1\s B\s\dots\s a\s n)\stackrel{n-1}\mra
(\dots\s n-1\cup B\s\dots\s a\s n) \\
\stackrel{a\cup}\sra (\dots\s  n-1\cup B\s\dots\s \{n,a\})
\end{multline}

Assume now $\sigma=(\dots\s n-1\cup A\s n)$.  Use the moves 
\eqref{eq:m21} and \eqref{eq:m22} to reduce it to the previously 
considered cases.
\begin{multline}\label{eq:m21}
(\dots\s n-1\cup A\s n)\stackrel{n-1}\mra
(\dots\s n-1\s A\s n) \\ \stackrel{A}\sra 
(\dots\s n-1\s x\s A\sm x\s n),\text{ where }|A|\geq 2.
\end{multline}
\begin{equation}\label{eq:m22}
(\dots\s \{n-1,a\}\s n)\stackrel{n-1}\mra
(\dots\s n-1\s a\s n) \stackrel{a\cup}\sra (\dots\s n-1\s \{n,a\})
\end{equation}

Finally assume $\sigma=(\dots\s B\s n-1\s A\s n)$. We consider three cases. 
If $|B|\geq 2$, we use the move \eqref{eq:m23}; if $|B|=1$, we use 
the move \eqref{eq:m24}; if $B=\emptyset$  use the move \eqref{eq:m25}, 
which is possible since $n\geq 3$. 
\begin{equation}\label{eq:m23}
(\dots\s B\s n-1\s A\s n)\stackrel{n-1}\mra
(\dots\s B\s  n-1\cup A\s n) \stackrel{B}\sra 
(\dots\s x\s B\sm x\s  n-1\cup A\s n)
\end{equation}
\begin{equation}\label{eq:m24}
(\dots\s b\s n-1\s A\s n)\stackrel{n-1}\mra
(\dots\s b\s  n-1\cup A\s n) \stackrel{b\cup}\sra 
(\dots\s \{n-1,b\}\cup A\s n)
\end{equation}
\begin{equation}\label{eq:m25}
(n-1\s [n-2]\s n)\stackrel{n-1}\mra
([n-1]\s n) \stackrel{n-2\s}\sra 
(n-2\s [n-3]\cup n-1\s n)
\end{equation}
In each case we reduce to the already considered case
$\sigma=(\dots\s A\cup n-1\s n)$.
\qed

%\begin{lm}\label{lm:2}
%Assume $\sigma$ and $\tau$ are two non-critical $n$-simplices of
%$\chi(\da^n)$ of opposite orientations, then there exists
%a~non-augmenting non-self-intersecting path $Q$ connecting $\sigma$
%with~$\tau$.
%\end{lm}
%\pr
%\qed

\begin{rem} Curiously, Lemma~\ref{lm:1} is not true when $n=2$ for
$\sigma=(1\s 0\s 2)$. It still remains true for all other simplices
  $\sigma$, and it is also trivially true for $n=1$. All of this can
  be seen by direct inspection.
\end{rem}

\begin{thm}\label{thm:cond1}
For any natural number $n$, the graph $\gr_n$ is conducting of the
first type.
\end{thm}
\pr The cases $n=1$ and $n=2$ can be verified directly, so we assume
$n\geq 3$. Recall now, that by Proposition~\ref{prop:sm} the standard
matching $M_\Sigma$ is a~near-perfect matching with a~critical
$n$-simplex indexed by~$\Sigma$.  Let $\tau$ be any other $n$-simplex
such that $\ori(\tau)=\ori(\Sigma)$. By Proposition~\ref{prop:bg}(1)
we need to show that there exists a~weakly semi-augmenting path
between $\tau$ and~$\Sigma$.  Let $\sigma$ be any of the neighboring
$n$-simplices of~$\Sigma$.  Since both $\sigma$ and $\tau$ are
non-critical, Lemma~\ref{lm:1} implies that there exist a~weakly
semi-augmenting path $Q_1$ between $\tau$ and $\{\rho,\nu\}$, and
a~weakly semi-augmenting path $Q_2$ between $\sigma$ and
$\{\rho,\nu\}$.  Since $\ori(\sigma)\neq\ori(\tau)$, the paths $Q_1$
and $Q_2$ will link them to the different vertices in
$\{\rho,\nu\}$. Let $Q$ be the concatenation of $Q_1$, followed by the
edge $(\rho,\nu)$, then by $Q_2$, and finally by the edge
$(\sigma.\Sigma)$. Clearly, $Q$ is a~semi-augmenting path from $\tau$
to~$\Sigma$.  \qed

\begin{thm}\label{thm:cond2}
Assume $[n]\supset V\neq\emptyset$, such that $|V|\leq n-2$, 
then $\gr_n(V)$ is conducting of the second type.
\end{thm}
\pr Note, that the assumption $n\geq 3$ is automatic here, since
$1\leq|V|\leq n-2$.  Without loss of generality, we can assume that
$V=[k]$, with $0\leq k\leq n-3$, and set $\Sigma:=\{k+1,\dots,n\}$.
By Proposition~\ref{prop:sm2}, the standard matching $M_\Sigma$ is
a~perfect matching on the graph $\gr_n(V)$. Let $\rho$ and $\nu$ be as
above, we still have $\rho,\nu\in\gr_n(V)$ and $\nu:=M_\Sigma(\rho)$.
Pick an arbitrary $\sigma\in\gr_n(V)$, which is non-critical with
respect to~$M_\Sigma$.  By Lemma~\ref{lm:1} there exists a~weakly
semi-augmenting path $Q$ from $\sigma$ to $\{\rho,\nu\}$.  All we need
to do is to see that this path lies entirely within $\gr_n(V)$.

Recall, that there are 4 types of edges used in the construction of
our path: $\stackrel{k}{\mra}$, $\stackrel{x\cup}{\sra}$,
$\stackrel{x\s}{\sra}$, and $\stackrel{A_i}{\sra}$. Clearly, if
$\sigma\in\Gamma_n(V)$, then any of the edges $\stackrel{k}{\mra}$,
$\stackrel{x\cup}{\sra}$, $\stackrel{A_i}{\sra}$ will lead to a~vertex
in $\Gamma_n(V)$ as well. The edges $\stackrel{x\s}{\sra}$ occur only
in paths \eqref{eq:m8}, \eqref{eq:m14}, \eqref{eq:m15}, and
\eqref{eq:m25}.  In all these cases we have $x\in\{n-2,n-1\}$, in
particular $x\notin V$; hence also here the edge
$\stackrel{x\s}{\sra}$ leads to a~vertex in~$\Gamma_n(V)$.

%Detailed analysis of our proof of (1) shows that all the operations
%from \eqref{eq:m1} to \eqref{eq:m25} which we used when constructing
%the path $Q$ will respect $\gr_n(V)$. Indeed, if $\sigma\in\gr_n(V)$,
%then $M_\Sigma(\sigma)\in\gr_n(V)$, so we only need to make sure that
%we stay within $\gr_n(V)$ when crossing the edges $\stackrel{a}\sra$.
%For such edges, the only problem which may occur would be if we split
%$A_1\s\dots\s A_t\stackrel{a}\sra a\s A_1\sm a\s\dots\s A_t$, under
%the conditions $A_1\not\subseteq V$ and $a\in V$. Since we never
%restrict ourselves in our moves which element is split off, we can
%always make sure that $a\notin V$, so this problem will not occur.

We conclude that the entire path stays within $\gr_n(V)$.  Now, we
have seen in the proof of (1) that the existence of such paths means
that any two vertices $\sigma,\tau$ of $\gr_n(V)$, such that
$\ori(\sigma)=\ori(\tau)$, are connected by an augmenting path.  By
Proposition~\ref{prop:bg}(2) this implies that $\gr_n(V)$ is
conducting of the second type.
\qed

\begin{thm}\label{thm:cond3} 
The graphs $\gr_5(V,\Omega)$ are conducting for all families of
prefixes $\Omega$ listed in Table~\ref{table:exc-simp}, under the
assumption $|S_p|\geq 3$.
\end{thm}
\pr Again without loss of generality we can assume that $n-2,n-1,n\in
S_p$, all other cases can be reduced to this one by appropriate
renaming. Choose $\Sigma$ to be any order on $[n]\sm V$ with
$n-2,n-1,n$ coming last, i.e., $\Sigma=(\dots,n-2,n-1,n)$.  By
Remark~\ref{rem:sn} the standard matching $M_\Sigma$ is a~well-defined
matching on the graph $\gr_n(V,\Omega)$ with one critical simplex for
each full prefix in $\Omega$. As above, we still have
$\rho,\nu\in\gr_n(V)$ and $\nu:=M_\Sigma(\rho)$.

Let $\sigma$ be a~non-critical simplex in $\gr_n(V,\Omega)$.  By
Lemma~\ref{lm:1} there exists a~weakly semi-augmenting path $Q$ from
$\sigma$ to $\{\rho,\nu\}$. This time, what we need to do is to see
that this path lies entirely within $\gr_n(V,\Omega)$. Again we need
to perform a~detailed analysis of moves \eqref{eq:m1}
through~\eqref{eq:m25}.

Take any edge $e=(\sigma,\tau)$ on our path.  Assume
$\sigma\in\Gamma_n(V,\Omega)$, say
$\sigma=(A_1,\dots,A_k,A_{k+1},\dots,A_t)$, where
$A_1,\dots,A_k\subseteq V$, and $A_{k+1}\not\subseteq V$. Of course we
have $(A_1\s\dots\s A_k)\in\Omega$. We need to show that
$\tau\in\Gamma_n(V,\Omega)$. We consider separately each of the 4
possible types of edge~$e$. 

\noindent
{\bf Case 1.} If $e$ is of the type $\stackrel{k}{\mra}$, then the
$\tau$ starts with the same prefix as $\sigma$, so the claim follows.

\noindent
{\bf Case 2.} Assume $e$ is of the type $\stackrel{A_i}{\sra}$.  If
$1\leq i\leq k$, then, since $\Omega$ is a~closed prefix, by
Definition~\ref{df:closed}(2) there exists $x\in A_i$, such that the
edge $\stackrel{x\s}{\sra}$ will lead to $\tau\in\Gamma_n(V,\Omega)$.
If $k+2\leq i\leq t$, then $\tau$ starts with the same prefix as
$\sigma$, so again $\tau\in\Gamma_n(V,\Omega)$. Finally, the case
$i=k+1$ never occurs on our path.

\noindent
{\bf Case 3.} Assume $e$ is of the type $\stackrel{x\s}{\sra}$.  As
mentioned in the proof of Theorem~\ref{thm:cond2}, in all of these
cases we have $x\in\{n-2,n-1\}$. This means that $x\notin V$, and
$\tau$ starts with the same prefix as $\sigma$, so
$\tau\in\Gamma_n(V,\Omega)$.

\noindent
{\bf Case 4.} Assume finally $e$ is of the type
$\stackrel{x\cup}{\sra}$. We must have $\{x\}=A_i$, for some $1\leq
i\leq t$. If $i\geq k+1$, then $\tau$ starts with the same prefix as
$\sigma$, so $\tau\in\Gamma_n(V,\Omega)$. If $i=k$, then $\tau$ starts
with the prefix $(A_1\s\dots\s A_{k-1})$. Since $\Omega$ is a closed
prefix, Definition~\ref{df:closed}(1) implies that $(A_1\s\dots\s
A_{k-1})\in\Omega$, hence $\tau\in\Gamma_n(V,\Omega)$. Finally, the 
case $i\leq k-1$ never occurs on our path.

%As before, if $\sigma\in\gr_n(V)$, then $M_\Sigma(\sigma)\in\gr_n(V)$,
%so we only need to make sure that we stay within $\gr_n(V)$ when
%crossing the edges $\stackrel{a}\sra$. Such an edge will denote either
%a~merging or a~splitting.
%
%Assume first that we have a~splitting $(\dots\s
%A_i\s\dots)\stackrel{a}\sra(\dots\s a\s A_i\sm a\s\dots)$.  If
%$A_j\not\subseteq V$ for some $j<i$, then obviously we stay within
%$\gr_n(V,\Omega)$. If $A_i\not\subseteq V$, then we can choose
%$a\notin V$, and again we stay within $\gr_n(V,\Omega)$. Finally, if
%$A_j\subseteq V$ for all $j\leq i$, then we are splitting a set within
%our prefix, and the claim follows from the fact that $\Omega$ is
%closed.%
%
%Assume now that we have a~merging $(\dots\s a\s
%A_i\s\dots)\stackrel{a}\sra(\dots\s  a\cup A_i\s\dots)$, where
%$A_{i-1}= a$. Obviously, the only problem would be the situation
%when $A_j\subseteq V$ for all $j\leq i-1$. If $A_i\not\subseteq V$,
%then we stay within $\gr_n(V,\Omega)$ because $\Omega$ is closed. If
%$A_i\subseteq V$, then we are in trouble, since we are merging two
%sets inside a prefix. However, the detailed analysis of our moves
%shows that this never happens.

Thus we have shown that $\tau\in\Gamma_n(V,\Omega)$ in all 4 cases.
If $\Omega$ has no full prefixes, then by Lemma~\ref{lm:sml}(1)
$M(\Sigma)$ is a~perfect matching and $\gr_n(V,\Omega)$ is conducting.
If $\Omega$ has a~single full prefix, then by Lemma~\ref{lm:sml}(2) it
has a near-perfect matching. Let $\gamma$ denote the critical
$n$-simplex. By an argument identical to the one used in the proof of
Lemma~\ref{lm:1}, we see that there is a~weakly semi-augmenting path
from $\gamma$ to any $n$-simplex $\sigma$ of $\gr_n(V,\Omega)$, such
that $\ori(\gamma)=\ori(\sigma)$. Using Proposition~\ref{prop:bg}(1),
we then conclude that $\gr_n(V,\Omega)$ is conducting.

Finally, when considering the row $(\{a,b,c\},\{d,e,f\})$, with
$a<b<c$, $d<e<f$, we get a~family of prefixes with 3 full prefixes,
hence we get 3 critical simplices. These simplices are
$\tau_1=(\{a,b\}\s c\s d\s e\s f)$, $\tau_2=(a\s b\s c\s d\s e\s f)$,
$\tau_3=(a\s \{b,c\}\s d\s e\s f)$.  We now extend our matching
$M_\Sigma$ by matching $\tau_2$ with $\tau_3$, obviously these share
an $(n-1)$-simplex, so the new matching is well-defined.  All that remains
is to find semi-augmenting paths from $\tau_2$ and $\tau_3$ to
$\{\rho,\nu\}$. To find such a path starting from $\tau_2$ we use the move
\eqref{eq:m27}, while to find such a path starting from $\tau_3$ we
use the move \eqref{eq:m28}.
\begin{equation}\label{eq:m27}
a\s b\s c\s d\s e\s f\mra
a\s \{b, c\}\s d\s e\s f\stackrel{e\cup}\sra 
a\s \{b, c\}\s d\s \{e, f\}
\end{equation}
\begin{equation}\label{eq:m28}
a\s \{b,c\}\s d\s e\s f\mra
a\s b\s c\s d\s e\s f\stackrel{e\cup}\sra 
a\s b\s c\s d\s \{e, f\}
\end{equation}
 In each case, we arrive at a~non-critical simplex, from which we have
 already found a~weakly semi-augmenting path to $\{\rho,\nu\}$.  \qed

\subsection{Explicit description of 21 paths}
\vs Let us now return to the labeling $I$, where we want to find
a~perfect matching on the set of all $1$-monochromatic
$5$-simplices. Using the notation of subsection~\ref{ss:3.4}, this set
is a~disjoint union of the sets $\cn(\sigma)$, where $\sigma$ ranges
over all $5$-simplices of $\chi(\da^5)$. The complete description of
$\cn(\sigma)$ has been obtained in Proposition~\ref{prop:3.11}.

At this point we would like to distinguish some of the $5$-simplices 
of~$\gr_5$. We set
\[\begin{array}{lcl}%\label{eq:cw}
\cw_1&:=&\{(a\s [5]\sm a)\s a\in[5]\}\\
\cw_2&:=&\{(\{a,b\}\s [5]\sm\{a,b\})\s a,b\in[5],\, a<b\}\\
\cw_3&:=&\{(\{a,b,c\}\s [5]\sm\{a,b,c\})\s a,b,c\in[5],\,a<b<c\}\\
\cw&:=&\cw_1\cup\cw_2\cup\cw_3\cup([5]).
\end{array}\]
Clearly, $|\cw|=42$, and we shall call the $5$-simplices in $\cw$ 
{\it exceptional simplices}.

For every $\sigma\in\gr_5$, $\sigma=(A_1\s\dots\s A_t)$, we now fix
a~matching on $\cn(\sigma)$. Specifically, we set
$\Sigma:=\st(\sigma)$, see Definition~\ref{df:st}, and let
$\wti\cm_\sigma$ to be the standard matching $\cm_\Sigma$, for all
$\sigma\notin\cw_3$. For $\sigma\in\cw_3$, say
$\sigma=(\{a,b,c\},\{d,e,f\})$, with $a<b<c$, $d<e<f$, we obtain
$\wti\cm_\sigma$ from the standard matching $\cm_\Sigma$, by extending
it by one more edge, matching $(a\s b\s c\s d\s e\s f)$ with $(a\s
\{b,c\}\s d\s e\s f)$, as in the proof of
Theorem~\ref{thm:cond3}.  We summarize what we know about these
matchings in the following proposition.

\begin{prop} \label{prop:42}
The matching $\wti\cm_\sigma$ is perfect for all $\sigma\notin\cw$,
and near-perfect for all $\sigma\in\cw$. 
\end{prop}
\pr If $\sigma=([5])$, Proposition~\ref{prop:sm} implies that $\wti\cm_\sigma$
is near-perfect. Assume now $\sigma\neq([5])$.
By Proposition~\ref{prop:sm2} the matching $\wti\cm_\sigma$ is perfect for all
$\sigma$ which do not appear in Table~\ref{table:exc-simp}. Assume finally 
that $\sigma$ does appear in Table~\ref{table:exc-simp}. A~line-by-line 
analysis shows that the corresponding prefix family $\Omega$ has full prefixes 
if and only if $\sigma\in\cw$. It has one full prefix if $\sigma\in\cw_1\cup\cw_2$,
and it has three full prefixes if $\sigma\in\cw_3$.
The result now follows from Lemma~\ref{lm:sml}, and the fact that two out
of three critical (with respect to the standard matching) simplices for 
$\sigma\in\cw_3$ have been matched in the extended matching~$\wti\cm_\sigma$. 
\myqed

We let $\wti\cm$ denote the matching on $\gr_5$ obtained as a union
of all matchings $\wti\cm_\sigma$. By Proposition~\ref{prop:42}
this matching has $42$ critical simplices. We connect these $42$ 
critical simplices by $21$ non-intersecting augmenting paths; 
the Table~\ref{table:21paths} contains the explicit description of 
the $21$ paths. 
This will allow to use the construction from Definition~\ref{df:mdef} 
and to deform the matching $\wti\cm$ to a~perfect matching~$\cm$.

\begin{table}[hbt]
\[\begin{array}{ccccccccc}
012&\ra& 0\s 12&\ra& 0\s 1\s 2&\ra& 0\s 1&\ra& 01\\
013&\ra& 0\s 13&\ra& 0\s 3\s 1&\ra& 0\s 3&\ra& 03\\
014&\ra& 0\s 14&\ra& 0\s 4\s 1&\ra& 0\s 4&\ra& 04\\
015&\ra& 0\s 15&\ra& 0\s 5\s 1&\ra& 0\s 5&\ra& 05\\
023&\ra& 0\s 23&\ra& 0\s 2\s 3&\ra& 0\s 2&\ra& 02\\
024&\ra& 4\s 02&\ra& 4\s 2\s 0&\ra& 4\s 2&\ra& 24\\
025&\ra& 2\s 05&\ra& 2\s 5\s 0&\ra& 2\s 5&\ra& 25\\
034&\ra& 3\s 04&\ra& 3\s 4\s 0&\ra& 3\s 4&\ra& 34\\
035&\ra& 3\s 05&\ra& 3\s 5\s 0&\ra& 3\s 5&\ra& 35\\
045&\ra& 4\s 05&\ra& 4\s 5\s 0&\ra& 4\s 5&\ra& 45\\
123&\ra& 1\s 23&\ra& 1\s 2\s 3&\ra& 1\s 2&\ra& 12\\
124&\ra& 1\s 24&\ra& 1\s 4\s 2&\ra& 1\s 4&\ra& 14\\
125&\ra& 1\s 25&\ra& 1\s 5\s 2&\ra& 1\s 5&\ra& 15\\
134&\ra& 3\s 14&\ra& 3\s 1\s 4&\ra& 3\s 1&\ra& 13\\
234&\ra& 2\s 34&\ra& 2\s 3\s 4&\ra& 2\s 3&\ra& 23\\

135&\ra& 1\s 35&\ra& 1\s 3\s 5&\ra& 1\s 3 &\ra&1\\
145&\ra& 4\s 15&\ra& 4\s 1\s 5&\ra& 4\s 1 &\ra&4\\
235&\ra& 3\s 25&\ra& 3\s 2\s 5&\ra& 3\s 2 &\ra&3\\
245&\ra& 2\s 45&\ra& 2\s 4\s 5&\ra& 2\s 4 &\ra&2\\
345&\ra& 5\s 34&\ra& 5\s 3\s 4&\ra& 5\s 3 &\ra&5
\end{array}\]

\[0\ra 012345\]
%\vspace{0.1cm}
\caption{Explicit description of the 21 paths.}
\label{table:21paths}
\end{table}

We define $\sig:\cw\ra\{\pm 1\}$, by setting
$\sig(\sigma):=\ori(\tau)$, where $\tau\in\gr_5^2$ is the unique
critical $5$-simplex in $\cn(\sigma)$.  Note that $\sig([5])=-1$,
$\sig(\sigma)=1$ for $\sigma\in\cw_1\cup\cw_2$, and $\sig(\sigma)=-1$
for $\sigma\in\cw_3$. We now need the following result.
\begin{lm}\label{lm:path}
Assume that we have a path~$Q$ in $\gr_5$ connecting two exceptional
$5$-simplices $\sigma$ and $\tau$, such that
\begin{enumerate}
\item[(1)] $\sig(\sigma)=-\sig(\tau)$;
\item[(2)] any $5$-simplex $\gamma$ on the path $Q$ other than $\sigma$
and $\tau$ is non-exceptional, and the corresponding graph $\cn(\gamma)$ 
is conducting of the second type.
\end{enumerate} 
Then there exists an augmenting path $T$ in $\gr_5^2$ such that
\begin{enumerate}
\item[(1)] $T$ connects the critical simplex in $\cn(\sigma)$ with the
  critical simplex in $\cn(\tau)$;
\item[(2)] $T$ lies entirely in $\bigcup_{\gamma\in Q}\cn(\gamma)$.
\end{enumerate}
\end{lm}
\pr Assume $Q=(\gamma_1,\dots,\gamma_t)$, with $\gamma_1=\sigma$,
$\gamma_t=\tau$. For all $1\leq i\leq t-1$, choose $5$-simplices
$\varphi_i\in\cn(\sigma_i)$, $\psi_i\in\cn(\sigma_{i+1})$ such that
\begin{enumerate}
\item[(1)] $\varphi_i$ and $\psi_i$ are adjacent in $\gr_5^2$;
\item[(2)] $\ori(\varphi_i)=\sig(\sigma)$.
\end{enumerate}
To see that such simplices can be chosen let us pick $\alpha$ and
$\beta$ to be two adjacent $5$-simplices in $\gr_5$, say
$\alpha=(A_1\s\dots\s A_p)$, $\beta=(B_1\s\dots\s B_q)$, and
$\beta=\cf(\alpha,x)$. Without loss of generality we can assume that
$A_p\subseteq B_q$; in fact, we could always assume that either
$A_p=B_q$, or $A_p\cup x =B_q$, but we do not need such detail. Pick
any $y\in A_p$.  Then any $5$-simplex $(\alpha\ds C_1\s\dots\s C_r\s
x)$ in $\gr_5^2$, such that $y\in C_1$ belongs to
$\cn(\alpha)$. Furthermore, it is adjacent to the $5$-simplex
$(\beta\ds C_1\s\dots\s C_r\s x)$, which in turn belongs to
$\cn(\beta)$.  There are $154$ such $5$-simplices, exactly half of
which have orientation $1$, so there is plenty of choice for the
simplices $\varphi_i$ and $\psi_i$.

Note that $\ori(\psi_i)=\sig(\tau)$ follows automatically from
$\ori(\varphi_i)=-\ori(\psi_i)$ and $\sig(\sigma)=-\sig(\tau)$.  Set
$\psi_0$ to be the critical $5$-simplex in $\cn(\sigma)$, and set
$\varphi_t$ to be the critical $5$-simplex in $\cn(\tau)$.  By
Theorem~\ref{thm:cond3} there exist semi-augmenting paths $Q_i$, for
$1\leq i\leq t$, such that each $Q_i$ connects $\psi_{i-1}$ with
$\varphi_i$. Concatenating these paths will yield the desired
path~$T$.  \myqed

Since the paths presented in Table~\ref{table:21paths} are
non-intersecting, it follows from Lemma~\ref{lm:path} that the
corresponding $21$ augmenting paths in $\gr_5^2$ are non-intersecting
as well. As we said above, this means that we can deform the matching
$\wti\cm$ to a~perfect matching $\cm$, which leads to the proof of our
main theorem.
Before we proceed with the proof, we need one last piece of terminology.

\begin{df}
Assume $v=(\sigma_1\ds\dots\ds\sigma_d)$ is a vertex of
$\chi^d(\da^n)$. The {\bf predecessor} of $v$ is the vertex of the
simplex $\tau=(\sigma_1\ds\dots\ds\sigma_{d-1})$ of
$\chi^{d-1}(\da^n)$, which has the same color as $v$; we shall denote
it by $\pred(v)$.
\end{df}

%\vskip5pt

\nin {\bf Proof of Theorem~\ref{thm:main-math}}.  Let us describe an
assignment $L$ of values $0$ and $1$ to vertices of
$\chi^3(\da^5)$. Let $w=(\sigma_1\ds\sigma_2\ds\sigma_3)$ be such
a~vertex, and set $\tau:=(\sigma_1\ds\sigma_2)$, $\tau$ is a~simplex
of $\chi^2(\da^5)$. If $\dim\tau\leq 3$, then we set
$L(w):=I(\pred(w))$. It is a~compliant labeling, since the initial
labeling $I$ was chosen to be compliant. We shall call such the
assignment $I(\pred(-))$ the {\it default value}.

Assume now $\dim\tau=4$. If there is only one $5$-simplex of
$\chi^2(\da^5)$ containing $\tau$, or if the two $5$-simplices
containing $\tau$ are not matched by $\cm$, then we let $L(w)$ be the
default value. Otherwise, we set $L(w):=0$.

Finally, consider the case $\dim\tau=5$. If $\tau$ is not
a~$1$-monochromatic $5$-simplex in $I$, then we let $L(w)$ be the
default value. Otherwise, let $\cf(\tau,c)$ be the $5$-simplex which
is matched to $\tau$ by $\cm$. We set 
\begin{equation}
\label{eq:star1}
L(w):=\begin{cases}
1, & \text{ if } c=C(\sigma_3);\\
0, & \text{ otherwise.}
\end{cases}
\end{equation}

We shall now see that the obtained assignment $L$ has no monochromatic
$5$-simplices. Let $\sigma=(\sigma_1\ds\sigma_2\ds\sigma_3)$ be
a~$5$-simplex of $\chi^3(\da^5)$, and let $\tau=(\sigma_1\ds\sigma_2)$
be the corresponding $5$-simplex of $\chi^2(\da^5)$
containing~$\sigma$. If $\tau$ is not $1$-monochromatic in $I$, then
by our construction, all values of $L$ on vertices of $\tau$ are the
default ones, which means that the vertices of $\sigma$ have the same
values under $L$ as vertices of $\tau$ of the same color under~$I$.
In particular, $\sigma$ is not monochromatic.

Assume now $\tau$ is $1$-monochromatic under $I$, and assume $\tau$ is
matched to $\cf(\tau,c)$ under~$\cm$. By our construction the vertex
of $\sigma$ which has color $c$ is assigned the value $1$, hence
$\sigma$ cannot be $0$-monochromatic. Let $d\in[n]$, such that $c\neq
d$, and $d$ is almost maximal with respect to~$\sigma_3$; such an
element exists due to Remark~\ref{rem:amax}. Let $w$ be the color of
$\sigma$ with the color~$d$. By our construction $L(w)=0$, hence
$\sigma$ is not $1$-monochromatic either.
\myqed

%As a~corollary of Lemma~\ref{lm:path}, we see that we can modify the
%near-perfect matching $\cm$ into a~perfect one, if we succeed to
%connect the 42 exceptional $5$-simplices in $\cw$ by $21$
%non-intersecting paths in $\mych(\da^5)$, making sure that each of the
%paths connects $n$-simplices with different values of~$\ori$.

%\subsection{Producing a perfect matching}

\section{The distributed protocol solving Weak Symmetry Breaking 
for $6$ processes in $3$ rounds}\label{sect:5}

In this section we describe the application of our main theorem in
theoretical distributed computing and present an explicite distributed
protocol solving Weak Symmetry Breaking for $6$ processes in $3$
rounds.  To start with, we record the information obtained in
Section~\ref{sect:4} in a~table, which we call \exsimp. Namely, for
each path from Table~\ref{table:21paths} use Lemma~\ref{lm:path} to
produce an~augmenting path $Q$ in $\gr_5^2$. For every simplex
$(\sigma\ds\tau)$ from $Q$ we add an entry $(\sigma,\tau,c)$ to
\exsimp, where $c$ is the color of the edge from $Q$ which is adjacent
to $(\sigma\ds\tau)$, and which does not belong to original matching
$\wti\cm$ (it does belong to the final matching $\cm$).

\begin{figure}
\begin{lstlisting}[numbers=none]
protocol WSB6 (input: `$p$`)
`{\bf write}`; `{\bf read}`; `{\bf write}`; `{\bf read}`;
assume the current view is `$(\sigma'\ds\tau')$`
if `$\car(\sigma')\neq[n]$` and `$(\sigma'\ds\tau')$` is not in Table `\ref{table:exv2}` then `{\bf decide 0}`
`{\bf write}`; `{\bf read}`;
assume the current view is `$(\sigma\ds\tau\ds\gamma)$`
if `$|\car(\gamma)|\leq n-1$` then `{\bf decide~1}`
if `$\car(\gamma)=[n]$` then `$\xi:=\tau$`
else assume `$\tau=(A_1\s\dots\s A_t\ds B_1\s\dots\s B_t)$`, set `$q:=[n]\setminus(B_1\cup\dots\cup B_t)$`
	if `$q\notin A_1\cup\dots\cup A_t$` then `{\bf decide~1}` 
	else pick `$1\leq k\leq t$` such that `$q\in A_k$`
	set `$\xi:=(A_1\s\dots\s A_{k-1}\s q\s A_k\sm q\s A_{k+1}\s\dots\s A_t)$`
if `$p=\match(\sigma,\xi)$` then `{\bf decide~1}` 
else `{\bf decide~0}`
	
function `\match` (input: `$\xi,\sigma$`) 
if `$(\xi,\sigma,c)$` is in `\exsimp` then return `$c$`
else return `$\lev(\xi,\st(\sigma))$` 
\end{lstlisting}
\caption{Protocol solving Weak Symmetry Breaking for 6 processes in 3 rounds.}
\label{figure:two:protocol}
\end{figure}

The formal protocol is given in Figure~\ref{figure:two:protocol}.  The
verbal description is as follows. Let $p$ be the id of the process
running the protocol.

\vskip5pt

\nin {\bf Step 1.} Execute two rounds of write-read sequence. Assume
that the view of the process $p$ after $2$ rounds is
$v=(\sigma\ds\tau)$. If $\car(\sigma)=[n]$, then proceed to Step~2.
Else we have $\car(\sigma)\neq[n]$, in which case we check if $v$ is
one of the exceptional second round views from
Definition~\ref{df:exv}, see Table~\ref{table:exv2}.  If it is an
exceptional second round view from that table then proceed to Step~2,
else {\bf decide~0} and stop.

\vskip5pt

\nin {\bf Step 2.} Execute one more round of write-read. Assume that
the view of the process $p$ after $3$ rounds is
$(\sigma\ds\tau\ds\gamma)$. 
\begin{itemize}
\item If $|\car(\gamma)|\leq n-1$, then {\bf decide~1} and stop.
\item If $\car(\gamma)=[n]$, set $\xi:=\tau$ and proceed to Step~3.
\item If $|\car(\gamma)|=n$, assume $\tau=(A_1\s\dots\s A_t\ds
  B_1\s\dots\s B_t)$. We have $|B_1\cup\dots\cup B_t|=n$, so there
  exists $q\in[n]$, $q\neq p$, such that $B_1\cup\dots\cup
  B_t=[n]\sm q$. If $A_1\cup\dots\cup A_t=[n]\sm q$, then {\bf
    decide~1} and stop. Else, there exists $1\leq k\leq t$, such that
  $q\in A_k$. Now, set \[\xi:=(A_1\s\dots\s
  A_{k-1}\s q\s A_k\sm q\s A_{k+1}\s\dots\s A_t).\]
\end{itemize}

\vskip5pt

\nin {\bf Step 3.} If $p=\match(\sigma,\xi)$, then {\bf decide~1}, else {\bf
  decide~0} and stop.

\vskip5pt

\nin
Here is the procedure $\matcht$, which has two ordered set partitions
$\sigma$, $\xi$ of $[n]$ as input and returns a value from $[n]$.

\vskip5pt

\nin {\tt Procedure} $\matcht$. 

\nin If $(\sigma,\xi,c)$ is in \exsimp, set $\text{ output}\,:=c$,
else set
\[\text{ output}\,:=\lev(\xi,\st(\sigma)).\] 

%\vskip5pt

We note that the views from Table~\ref{table:exv2} can also be
described verbally. For example, the view
\[\begin{array}{|c||c|}
\hline
a & a \\ \hline
a & a \\ 
\hline
\end{array}\]
means that process $a$ has only seen itself both in the first and in
the second round. The view 
\[\begin{array}{|c||c|}
\hline
a,b,c & a,b \\ \hline
a,b & a \\ 
\hline
\end{array}\]
has a~more lengthy interpretation. It means, that the process $a$ has
seen one other process $b$ in the second round, whose id was larger
than that of~$a$. Furthermore, $a$ has seen 3 processes in the first
round, including itself and $b$. The third process it has seen has id
larger than that of both $a$ and $b$. Finally, the views of $b$ and
$a$ in the first round were the same ($a$ knows that after the second
round). We leave it to the reader to provide similar interpretations
for other views listed in the Table~\ref{table:exv2}.

\section{Weak Symmetry Breaking cannot be solved in $1$ round}\label{sect:6}

Currently, there are no lower bounds for the number of rounds needed
to solve the Weak Symmetry Breaking task. Here we give the first such
lower bound, stating that if Weak Symmetry Breaking can be solved at
all, one would need at least two rounds to do that.

\begin{thm}\label{thm:sbg2}
We have $\msb(n)\geq 2$, for all $n\geq 2$. In other words, the Weak
Symmetry Breaking task cannot be solved in one round for any value
of~$n\geq 2$.
\end{thm}
\pr Let $\cv^n$ denote the set of pairs of integers $\{(r,p)\s 1\leq
p\leq r\leq n\}$. Let us assume that we have a distributed protocol
running exactly one round and having output in the set
$\{0,1\}$. After the execution of this protocol each process $\rho$
has seen a certain number of other processes and has determined the
relative position of its own id among the id's of the processes it has
seen. Such an information can be encoded by a~pair of integers
$(r,p)$, such that $1\leq r\leq n$, this is the number of processes,
including itself, which $\rho$ has seen, and $1\leq p\leq r$, which is
the relative position of the id of $\rho$. Hence, the decision
function for such a protocol is precisely a function
$\delta:\cv^n\ra\{0,1\}$.

Assume now we are given such a decision function $\delta$ solving the
Weak Symmetry Breaking. Let us break up
$\cv^n=\cv^n_1\cup\dots\cup\cv^n_n$, where $\cv^n_i=\{(i,p)\s 1\leq
p\leq i\}$.  Let $k$ be the maximum index such that $\delta_{\cv^n_i}$
is not surjective.  Since $\delta_{\cv^n_1}$ is not surjective, the
number $k$ is well-defined.  Without loss of generality we can assume
that $\delta(k,1)=\delta(k,2)=\dots=\delta(k,k)=0$. For all $k+1\leq
i\leq n$ we let $\alpha_i$ be any number such that
$\delta(i,\alpha_i)=0$. Such a~number exists for all $k+1\leq i\leq
n$, since by our construction $\delta_{\cv_i}$ is surjective for these 
values of~$i$.

We shall now construct an execution $E$ of the protocol after which
the decision function will assign the value $0$ to all processes,
which of course contradicts to the claim that the protocol solves Weak
Symmetry Breaking. The execution $E$ will start with a~simultaneous
activation of $k$ processes, after which the rest of the processes
will activate one at a~time. In total the execution $E$ has $n-k+1$
rounds.

More specifically, we can find numbers $\xi(1),\dots,\xi(n)$, such
that the execution $E$ is described as follows:
\begin{enumerate}
\item[(1)] the id's of the processes participating in the first round
  are $\xi(1)$, $\dots$, $\xi(k)$;
\item[(2)] for each $r=2,\dots,n-k+1$, the id of the process participating
in the round $r$ is $\xi(k+r-1)$.
\end{enumerate} 

These numbers can be determined constructively as follows. To start
with, the numbers $\xi(1)$, $\dots$, $\xi(n)$ are initialized to
be~$0$. We then proceed in $n-k+1$ steps. In the first step, we set
$\xi(i):=i$ for all $1\leq i\leq k$. Next, for $i$ running from $k+1$
to $n$, repeat the following: set $\xi(i):=\alpha_i$, and increase by
$1$ each $\xi(j)$, such that $j<i$ and $\xi(j)\geq \alpha_i$. The
result after $n-k+1$ steps is then taken as the final output. 

It easy to see that each process activated in round $r$, for $2\leq
r\leq n-k+1$, will see $k+r-1$ processes, including itself, and that
the relative position of his id in what he sees will be
$\alpha_{k+r-1}$.  \qed

%\newpage

%\newpage

%$\,$

%\hfill

%\newpage

\end{document}